\documentclass[aps,prb,preprint,superscriptaddress,colorlinks=true,linkcolor=black,urlcolor=black,citecolor=black,anchorcolor=black,colorlinks,bookmarks=false,citecolor=blue,linkcolor=black,urlcolor=blue]{revtex4-1}

\usepackage{ulem}
\usepackage{bm,color,amsmath,amssymb,latexsym,graphicx,psfrag,accents,float}
\usepackage{amscd,dsfont,wasysym,mathrsfs,mathtools}
\usepackage{multirow}
\usepackage{dcolumn}
\usepackage{xcolor}
\usepackage{comment}
\usepackage{booktabs} 
\usepackage{upgreek}
\usepackage[utf8]{inputenc}
\usepackage{braket}
\usepackage{times}
\usepackage{amsfonts}
\usepackage{mathrsfs}
\usepackage{graphicx}
\usepackage{dcolumn}
\usepackage{bm}
\usepackage{color}

\usepackage[colorlinks,bookmarks=false,citecolor=blue,linkcolor=red,urlcolor=blue]{hyperref}
\newcolumntype{L}[1]{>{\raggedright\arraybackslash}p{#1}}
\newcolumntype{C}[1]{>{\centering\arraybackslash}p{#1}}
\newcolumntype{R}[1]{>{\raggedleft\arraybackslash}p{#1}}

\newcommand{\bpm}{\begin{pmatrix}}
\newcommand{\epm}{\end{pmatrix}}

\newcommand{\bal}{\begin{align}}

\usepackage{amsmath}
\usepackage{ulem}
\begin{document}
\renewcommand{\theequation}{S\arabic{equation}}
	\newcommand{\beginMethods}{%
		\setcounter{table}{0}
		\renewcommand{\thetable}{S\arabic{table}}%
		\setcounter{figure}{0}
		\renewcommand{\thefigure}{S\arabic{figure}}%
	}

\newcommand{\rim}[1]{\textcolor{green}{#1}}
\newcommand{\CVS}{CsV$_3$Sb$_5$ }

\title{Long-range electron coherence in Kagome metals}

\author{Chunyu (Mark) Guo${}^{\dagger}$}\affiliation{Max Planck Institute for the Structure and Dynamics of Matter, Hamburg, Germany}
\author{Kaize Wang${}^{}$}
\affiliation{Max Planck Institute for the Structure and Dynamics of Matter, Hamburg, Germany}
\author{Ling Zhang${}^{}$}
\affiliation{Max Planck Institute for the Structure and Dynamics of Matter, Hamburg, Germany}
\author{Carsten Putzke${}^{}$}
\affiliation{Max Planck Institute for the Structure and Dynamics of Matter, Hamburg, Germany}
\author{Dong Chen}\affiliation{Max Planck Institute for Chemical Physics of Solids, Dresden, Germany}\affiliation{College of Physics, Qingdao University, Qingdao, China}
\author{Maarten R. van Delft}\affiliation{High Field Magnet Laboratory (HFML - EMFL), Radboud University, Toernooiveld 7, 6525 ED Nijmegen, The Netherlands}
\affiliation{Radboud University, Institute for Molecules and Materials, Nijmegen 6525 AJ, Netherlands}
\author{Steffen Wiedmann}\affiliation{High Field Magnet Laboratory (HFML - EMFL), Radboud University, Toernooiveld 7, 6525 ED Nijmegen, The Netherlands}
\affiliation{Radboud University, Institute for Molecules and Materials, Nijmegen 6525 AJ, Netherlands}
\author{Fedor F. Balakirev}\affiliation{National High Magnetic Field Laboratory, Los Alamos National Laboratory, Los Alamos, New Mexico 87545, USA}
\author{Ross  D. McDonald}\affiliation{National High Magnetic Field Laboratory, Los Alamos National Laboratory, Los Alamos, New Mexico 87545, USA}
\author{Martin Gutierrez-Amigo}\affiliation{Department of Applied Physics, Aalto University School of Science, FI-00076 Aalto, Finland}
\author{Manex Alkorta}\affiliation{Centro de Física de Materiales (CSIC-UPV/EHU), Donostia-San Sebastian, Spain}
\affiliation{Department of Physics, University of the Basque Country (UPV/EHU), Bilbao, Spain}
\author{Ion Errea}\affiliation{Centro de Física de Materiales (CSIC-UPV/EHU), Donostia-San Sebastian, Spain}
\affiliation{Donostia International Physics Center, Donostia-San Sebastian, Spain}
\affiliation{Fisika Aplikatua Saila, Gipuzkoako Ingeniaritza Eskola, University of the Basque Country (UPV/EHU), Donostia-San Sebastian, Spain}
\author{Maia G. Vergniory}\affiliation{Donostia International Physics Center, Donostia-San Sebastian, Spain}
\affiliation{Max Planck Institute for Chemical Physics of Solids, Dresden, Germany}
\author{Takashi Oka}\affiliation{The Institute for Solid State Physics, The University of Tokyo, Kashiwa, Japan}
\author{Roderich Moessner}\affiliation{Max Planck Institute for the Physics of Complex Systems, Dresden, Germany}
\author{Mark H. Fischer}\affiliation{Department of Physics, University of Zürich, Zürich, Switzerland}
\author{Titus Neupert}\affiliation{Department of Physics, University of Zürich, Zürich, Switzerland}
\author{Claudia Felser}\affiliation{Max Planck Institute for Chemical Physics of Solids, Dresden, Germany}
\author{Philip J. W. Moll${}^{\dagger}$}\affiliation{Max Planck Institute for the Structure and Dynamics of Matter, Hamburg, Germany}

\date{\today}
\maketitle
\normalsize{$^\dagger$Corresponding authors: chunyu.guo@mpsd.mpg.de(C.G.); philip.moll@mpsd.mpg.de(P.J.W.M.).}

\section*{ABSTRACT}
{\bf The wave-like nature of electrons lies at the core of quantum mechanics, distinguishing them from classical particles. Landmark experiments have revealed phase coherence of mobile electrons within solids, such as Aharonov-Bohm interference in mesoscopic rings. However, this coherence is typically limited by numerous environmental interactions. Controlling and ideally mitigating such decoherence remains a central challenge in condensed matter physics. Here, we report magnetoresistance oscillations in mesoscopic pillars of the Kagome metal CsV$_3$Sb$_5$ for fields applied parallel to the Kagome planes. Their periodicity is independent of materials parameters, simply given by the number of flux quanta $h/e$ threading between adjacent Kagome layers akin to an atomic-scale Aharonov-Bohm interferometer. Intriguingly they occur under conditions not favorable for typical interference in solids, at temperatures above 20 K and in micrometer-scale devices well exceeding the single-particle mean free path. Further, the oscillations exhibit non-analytic field-angle dependence and scale consistently with a broad range of key electronic responses in CsV$_3$Sb$_5$, pointing to a cooperative mechanism that establishes intrinsic coherence. Our findings provide new insights into the debated origin of correlated order in CsV$_3$Sb$_5$ and establish Kagome metals as a promising platform for interaction-stabilized long-range electron coherence — crucial for both fundamental studies and technological advancements in quantum interference in metallic systems.}

\subsection*{Introduction}

A cornerstone of quantum mechanics is the recognition that electrons, unlike classical particles, are characterized not only by amplitudes but also by phases, enabling wave-like interference. This fundamental insight, first demonstrated by Davisson and Germer in 1927 in free-space electron beams\cite{davisson1927scattering}, extends to electrons in solids despite the severe suppression of coherence by environmental interactions. Inelastic processes such as electron-electron and electron-phonon scattering typically restrict such interference phenomena to sub-Kelvin temperatures\cite{olariu1985quantum}. Landmark experiments such as those by Webb \textit{et al.} detecting the Aharonov-Bohm effect in metallic rings showcased how microscopic structures can act as beam splitters, illustrating the detection and manipulation of quantum phases in metals\cite{washburn1986aharonov}. Today, the ability to probe, control, and utilize such phases underpins advancements in quantum technologies. This prompts a pivotal question about the material properties required to sustain and enhance long-range phase coherence in metals despite their complex internal dynamics. Quantum materials may offer a promising path by providing tailored environments that enhance coherence, opening new avenues for fundamental exploration and technological innovation. 

A notable case of electron interference in layered and highly anisotropic metals had been reported in (Pd,Pt)CoO$_2$\cite{putzke2020h}. Micrometer-sized rectangular bars along the out-of-plane direction were subjected to in-plane magnetic fields, leading to peculiar field-linear magnetoresistance oscillations. Their period is given by a magnetic flux quantum, $h/e$, threading between neighboring metallic (Pd,Pt)-planes which points to a mesoscopic interference phenomenon (Fig. 1a). Generally, interference patterns arise from constraining the possible paths of a wavepacket, as in the double-slit experiment or mesoscopic rings. Indeed, the period matches the intuitive expectation of an Aharonov-Bohm effect in a quantum ring formed by two adjacent layers in the stack connected at the opposite sidewalls, supported in essence by a more refined transport calculation via Kubo-formalism reproducing such oscillations in confined out-of-plane stacks. While the microscopic origin of coherence at high temperature ($\sim$ 60 K) and over long distances ($\sim 8~\mathrm{\mu}$m) remains unclear, it has been speculated to be rooted in the extraordinary purity of these materials exhibiting record mean free paths for oxides of $\sim 25~\mu$m at 2 K. Accordingly, micron-sized pillars readily enter the ballistic regime in which scattering is only dominated by the boundaries\cite{bachmann2022directional,bachmann2019super,nandi2018unconventional,moll2016evidence}, a natural setting for textbook quantum mechanics to emerge.

Here, we present the first observation of $h/e$ oscillations beyond (Pt,Pd)CoO$_2$ in a most unexpected place, the layered Kagome superconductor CsV$_3$Sb$_5$ ($T_{\text{c}}\sim2.8$ K). Its V-based Kagome layers take the role of the conductive planes despite its substantially less anisotropic conduction compared to PdCoO$_2$. Kagome physics manifests in \CVS through features in the electronic structure such as Dirac points and van Hove singularities, the essential ingredients of geometric orbital frustration needed to host coupled electronic orders. Indeed, its charge-order at $T_{\text{CDW}} \sim 94$~K\cite{ortiz2019new,ortiz2020cs,epc,gutierrez2024phonon} is anti-correlated with $T_{\text{c}}$ and its superconductivity has been proposed to be unconventional by some, but not all works\cite{Sbcris,gupta2022microscopic,liang2021three,zhang2023nodeless,zhao2024ultralow,le2024superconducting,hossain2024unconventional}. In addition, various experimental probes detect a thus-far enigmatic change of the electronic ground state in CsV$_3$Sb$_5$ at an intermediate temperature scale $T'\sim 30$~K, yet the microscopic nature of the low-temperature phase and broken symmetries (if any) remain under active debate\cite{Guo2022,xiang2021twofold,guo2024correlated,DresdenThermal,yu2021evidence,Nie2022,wei2024three,KagomeReview,Neupert2022}. In one leading hypothesis, this exotic state hosts persistent orbital currents, subsumed as orbital loop current circulating the Kagome plaquettes\cite{Christensen2,TitusAdd,Tazai,Denner,deng2024chiral,scammell2023chiral,ingham2025vestigial}.

The unexpected emergence of $h/e$ oscillations in \CVS is remarkable for two main reasons. First, the oscillations evidence the ability of the electronic fluid to interfere over distances well exceeding the transport or quantum mean free paths, challenging pictures of single-particle coherence. Second, the onset temperature of the oscillations strikingly coincides with $T'$, suggestive of their common origin and the existential role of a correlated many-body state in this long-range coherence.

\subsection*{Observation of $h/e$ oscillations in CsV$_3$Sb$_5$}

\begin{figure}
	\centering
        \includegraphics[width = 0.95\linewidth]{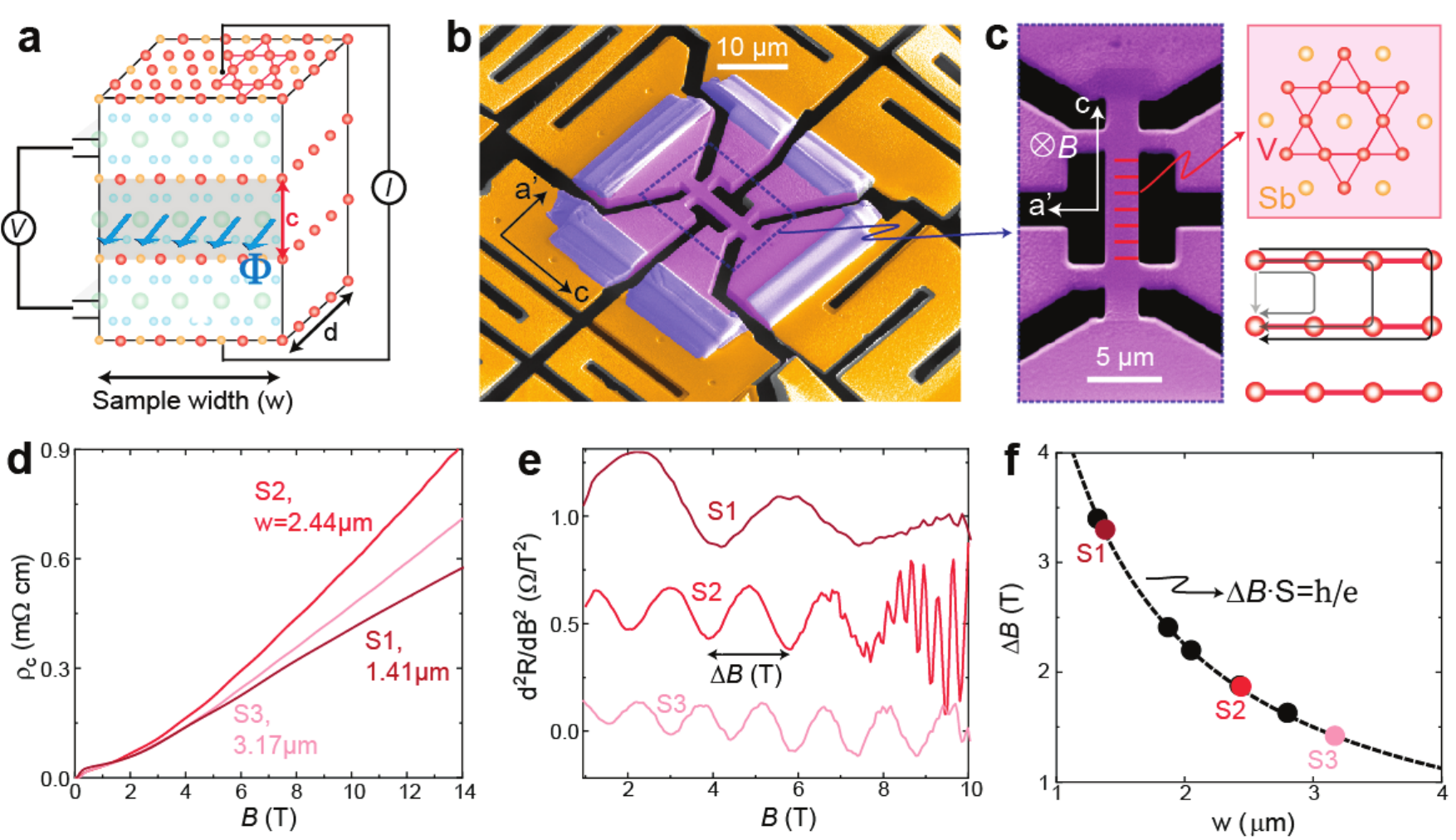}
		\caption{\textbf{$h/e$ oscillations in CsV$_3$Sb$_5$.} (a) Illustration of out-of-plane magnetotransport measurement setup for $h/e$ oscillations. The quantum flux box is constructed by the adjacent Kagome planes. The flux-entering area size, S, is defined by the width of the device and the lattice constant between the nearest Kagome layers. (b) Scanning electron imaging of the CsV$_3$Sb$_5$ microstructure hosted by soft membrane springs. (c) The zoom-in image of the device at the right hand clarifies the electrical configuration for out-of-plane resistance measurements as the current is applied perpendicular to the Kagome planes. On the left, we illustrate the Kagome layer made of Vanadium atoms and the possible quantum paths for electron hopping between two adjacent Kagome planes. (d) Field dependence of magnetoresistance of devices S1, S2, and S3. (e) The second derivative clearly reveals $h/e$ oscillations, the periodicity of which depends inversely on the sample width. (f) The relation between the oscillation period and device width follows the description of the magnetic flux enclosed by the Kagome layers. Periods measured in further samples are shown, see supplement for details.}
	\label{Intro}
\end{figure}

At the core of our experiment is the detection of field-periodic oscillations in the interlayer magneto-conduction of micron-sized pillars (Fig. \ref{Intro}), which can be viewed as microscopic stacks of Kagome planes. An electric current flows along the wire perpendicular to the Kagome layers while a transverse magnetic field $B$ aligned with the layers is applied. Its magnetoresistance (MR) at 2 K above the critical field of superconductivity $H_{\text{c2}}$ increases super-linearly and is superimposed by clearly observable periodic-in-field oscillations. These continue with increasing fields, though they merge into $1/B$-periodic Shubnikov-de-Haas (SdH) oscillations preventing their distinct identification at high magnetic fields.  Unlike SdH oscillations that encode a material's Fermiology and chemical potential, the period $\Delta B$ of this process is remarkably independent of material properties and uniquely determined by the magnetic field required to thread a flux quantum between two adjacent Kagome planes in the stack, $\Delta B \cdot w\cdot c=h/e \coloneqq \Phi_0$, combining the atomic interlayer distance ($c \approx$ 9 $\textup{\AA}$) and the macroscopic width of the bar $w$, which is fully controlled during fabrication and here typically chosen to fall in the 1 to 3 $\mu$m range. Varying the width systematically between devices allows us to unambiguously identify this universal scaling relation (Fig. \ref{Intro}f, note the dashed line is fitting-parameter-free).

Observing this effect in CsV$_3$Sb$_5$ relies on two recent methodological advances. First, to observe a measurable signal, macroscopic numbers of identical Kagome layers must be averaged. To that end, we employ Focused Ion Beam (FIB) machining to carve $c$-direction aligned bars from the plate-like single crystals. Second, CsV$_3$Sb$_5$ has been shown to be extraordinarily strain-sensitive\cite{Guo2022,guo2024correlated,guo2024distinct,Strain}. Hence, these delicate structures must be mechanically decoupled from any unintentional strain arising from thermal mismatch to the substrate\cite{Maja,Maarten}. This is achieved by suspending the samples on soft SiN$_x$ membranes coated with Au for electrical connection, as demonstrated elsewhere\cite{Guo2022,guo2024correlated,guo2024distinct}. Indeed, in substrate-mounted devices at typical strain levels for condensed matter experiments the oscillations are washed out (see supplement).

\begin{figure}
	\centering
        \includegraphics[width = 0.95\linewidth]{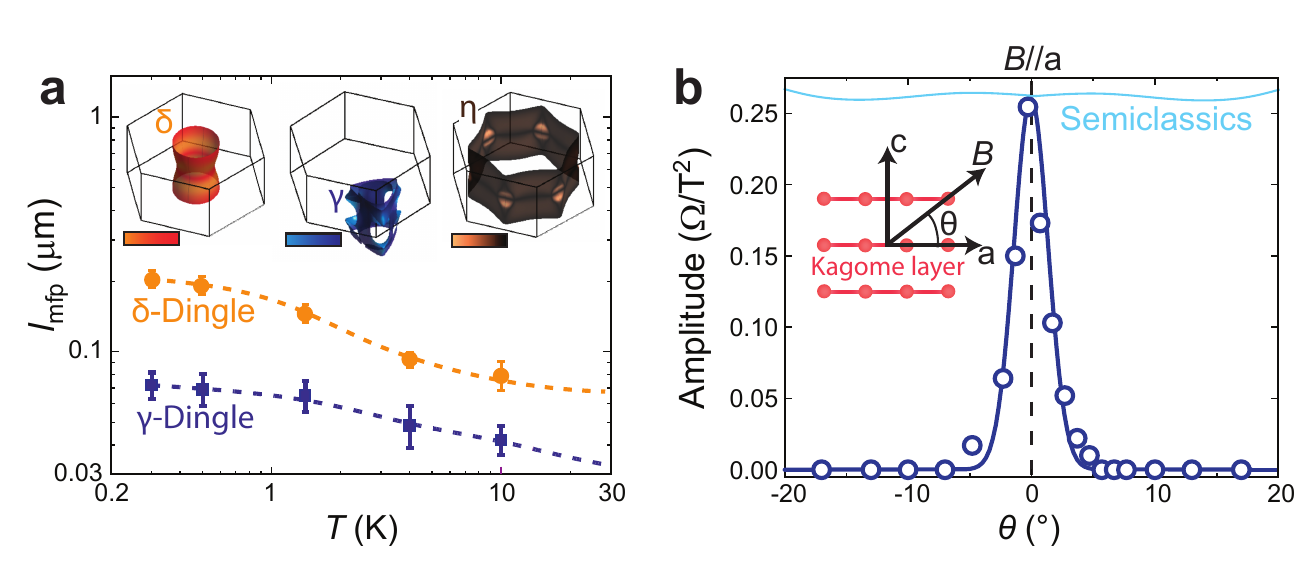}
		\caption{\textbf{Mean free path analysis and angular dependence of oscillation amplitude.} (a) Quantum mean free path extracted from the field dependence of the SdH oscillation amplitude based on Dingle analysis. The insets present different branches of Fermi surfaces in the Brillouin zone, and the color bar stands for the distribution of Fermi velocity anisotropy ($v_F^{ip}$/$v_F^z$) (b) Angular dependence of $h/e$ oscillation amplitude. The rotation of the field from in-plane to out-of-plane significantly reduces the oscillation amplitude. It vanishes when the angle $\theta$, defined as the angle between the magnetic field and the crystalline a-direction, exceeds $5^{\circ}$ (see also supplement).}
	\label{Out}
\end{figure}

The central characteristic is the mean free path, which if it greatly exceeds the device size, leads to ballistic electronic behavior. Such unperturbed electronic motion is a natural mechanism through which electrons obtain information about the mesoscopic size of their containment as the $w$-dependence of the periodicity unequivocally demonstrates. Intriguingly, however, both the quantum mean free path extracted by a Dingle analysis (Fig. \ref{Out}) as well as estimates of the transport mean free path from the in-plane resistivity are substantially smaller than any studied device, especially at high temperatures ($w_{max}$ = 3.2 $\mu$m see supplement for analysis). The quantum mean free paths ($l_q$) of all Fermi pockets observed by SdH oscillations fall below 200 nm above 1 K. Also the most optimistic estimates for the transport mean free path ($l_t \approx 500$ nm at $T_c$, see supplement) is at least six times smaller than the largest devices, and collapses further as the temperature is increased above 10~K. Such short transport mean free paths in CsV$_3$Sb$_5$ are not surprising, considering the formation of charge order domains\cite{liang2021three,zhao2021cascade}, the large phase-space for scattering in its complex multiband Fermiology\cite{huang2022mixed,ortiz2021fermi,chen2023magnetic,chapai2023magnetic}, the strong electron-phonon coupling as well as electronic interactions\cite{xie2022electron,he2024anharmonic,hu2022rich,kang2022}. Given the high zero-field resistance and the suppressed quantum mean free path one would not expect particularly coherent transport in the system.

One natural resolution may be that the oscillations do not require coherence. Indeed, a semiclassical ballistic theory has been proposed to explain the oscillations in PdCoO$_2$. In this Bloch-Lorentz model\cite{vilkelis2023bloch}, incoherent but ballistic quasiparticles slide on open orbits on its cylindrical Fermi surface, leading to commensurability oscillations and semi-classical localization periodic in the in-plane magnetic field. The high disorder levels of \CVS, however, are strongly detrimental to this mechanism.

\subsection*{Exotic angular dependence}

Further evidence for behavior not captured by semi-classics comes from the sharp suppression of the oscillations when the magnetic field is tilted away from ideal in-plane alignment towards the out-of-plane direction, where it becomes undetectable at angles above $5^{\circ}$ (Fig. \ref{Out}b). This strong reaction to field misalignment is remarkable for multiple reasons. First, the angle dependence mirrors quantitatively the recently reported suppression of the field-induced diode effect\cite{Guo2022} associated with field-tuning of its chiral electronic order (see supplement). As the oscillations continue into the zero-field limit, it implies a negligibly small threshold field along the out-of-plane direction suppresses the coherence. This low field response is further in qualitative agreement with recent optical experiments reporting chirality manipulation \cite{xing2024optical}. Second, the amplitudes vanish faster than expected for single-particle interference which only depends on the interlayer flux, slowly varying with misalignment angle $\theta$ as $\Phi(\theta)=B w c \cos(\theta)$. Equally, integrating the semi-classical Bloch-Lorentz equations on the cylindrical Fermi surfaces of \CVS yields a much weaker angle dependence.

\begin{figure}
	\centering
        \includegraphics[width = 0.95\linewidth]{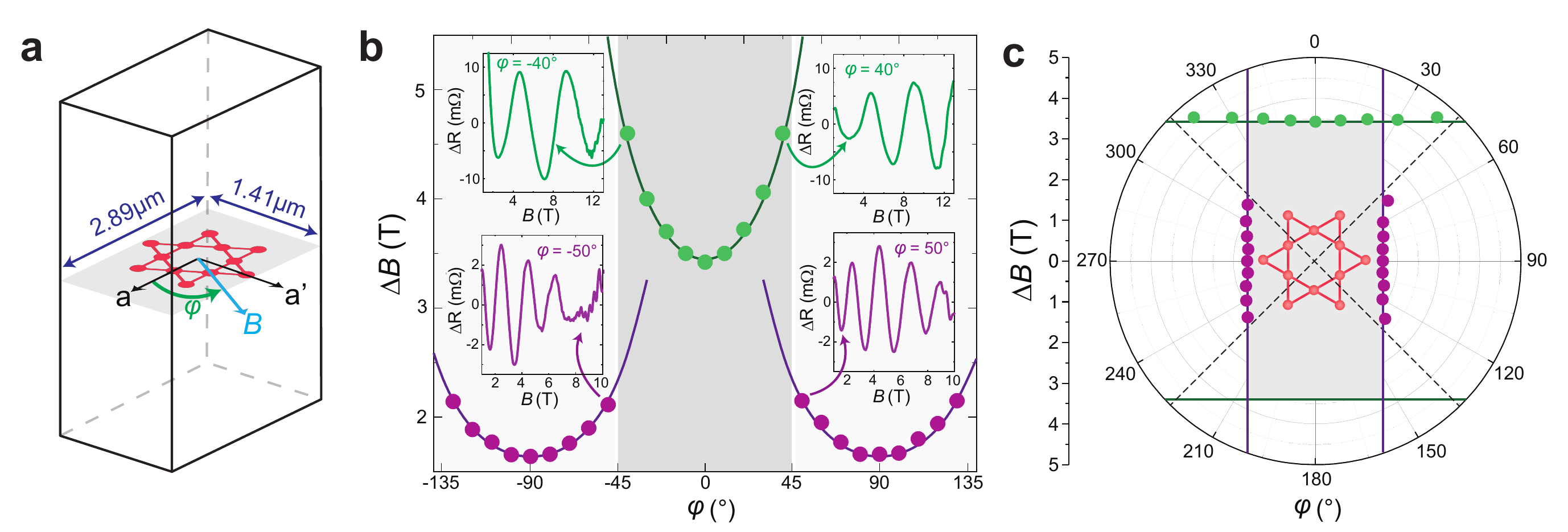}
		\caption{\textbf{Angular dependence: In-plane rotation.} (a) Illustration of field rotation within the Kagome plane. The angle $\varphi$ is defined as the angle between the magnetic field and crystalline a-direction. (b) Angular dependence of oscillation period in device S1 with magnetic field rotated within the Kagome plane. The configuration of in-plane rotation and its correspondence to the device's cross-section are illustrated in the left-hand inset. The distinct oscillation period with the field perpendicular or parallel to the a-direction projects the difference between its width (1.41$\mu m$) and depth (2.89$\mu m$). Most importantly, two abrupt changes in the oscillation period can be found at both -45 and 45 degrees. This is further elaborated by the field-dependent oscillations measured at both $\pm$40 and $\pm$50 degrees, as presented in the insets. The striking distinction between them further emphasizes the distinct switchings of oscillation frequency at all angles of 45 modulo 90 degrees. (c) The reconstructed angular dependence of the oscillation period is displayed on a polar plot. This further confirms that the distinct switching is unrelated to the device's geometric aspect ratio.}
	\label{In}
\end{figure}

Next, we demonstrate the unique behavior of the $h/e$ oscillations in CsV$_3$Sb$_5$ upon rotating the magnetic field in the plane (Fig. \ref{In}). Tilting the field away from the surface normal ($\varphi=0^{\circ}$) yields an increase of the period following a $\cos(\varphi)^{-1}$ scaling. This indicates that the quantum process remains tuned by the effective flux threading the surface, and the role of rotation merely reduces it as $\Phi(\varphi)=\Phi(0)\cos(\varphi)$. This trend abruptly terminates at $45^{\circ}$, where the period discontinuously doubles within this sample. Further increasing the angle towards the surface normal of the other sidewalls ($\varphi=90^{\circ}$) uncovers yet another $\cos(\varphi)^{-1}$ scaling, but now the effective width corresponds to this shorter sidewall (green branch in Fig. \ref{In}). Most strikingly, the switching angle of $45^{\circ}$ neither corresponds to any symmetry of the Kagome plane nor the rectangular shape of the device. Instead, the system selecting the surface most aligned with the field was consistently observed in all measured devices, as the switching was always found to occur at  $45^{\circ}$ irrespective of the geometric aspect ratio of the bar (see supplement also).

This further demonstrates a key difference to PdCoO$_2$, which hosts three one-dimensional electronic subsystems\cite{putzke2020h}. Each system oscillates independently, leading to a smooth angle evolution of three beating periods at any in-plane angle. The sharp jump between frequencies in CsV$_3$Sb$_5$ (see inset of Fig. 3a) clearly departs from this behavior, further mounting evidence against a description in terms of semi-classical orbits given the lack of any $45^{\circ}$ symmetry of the Fermi surface. On the contrary, it again demonstrates a non-analytical response to a magnetic field of the electronic system in CsV$_3$Sb$_5$, in accordance with other experimental probes\cite{Guo2022}.

\subsection*{Temperature dependence and global scaling at $T'$}

\begin{figure}
	\centering
        \includegraphics[width = 0.95\linewidth]{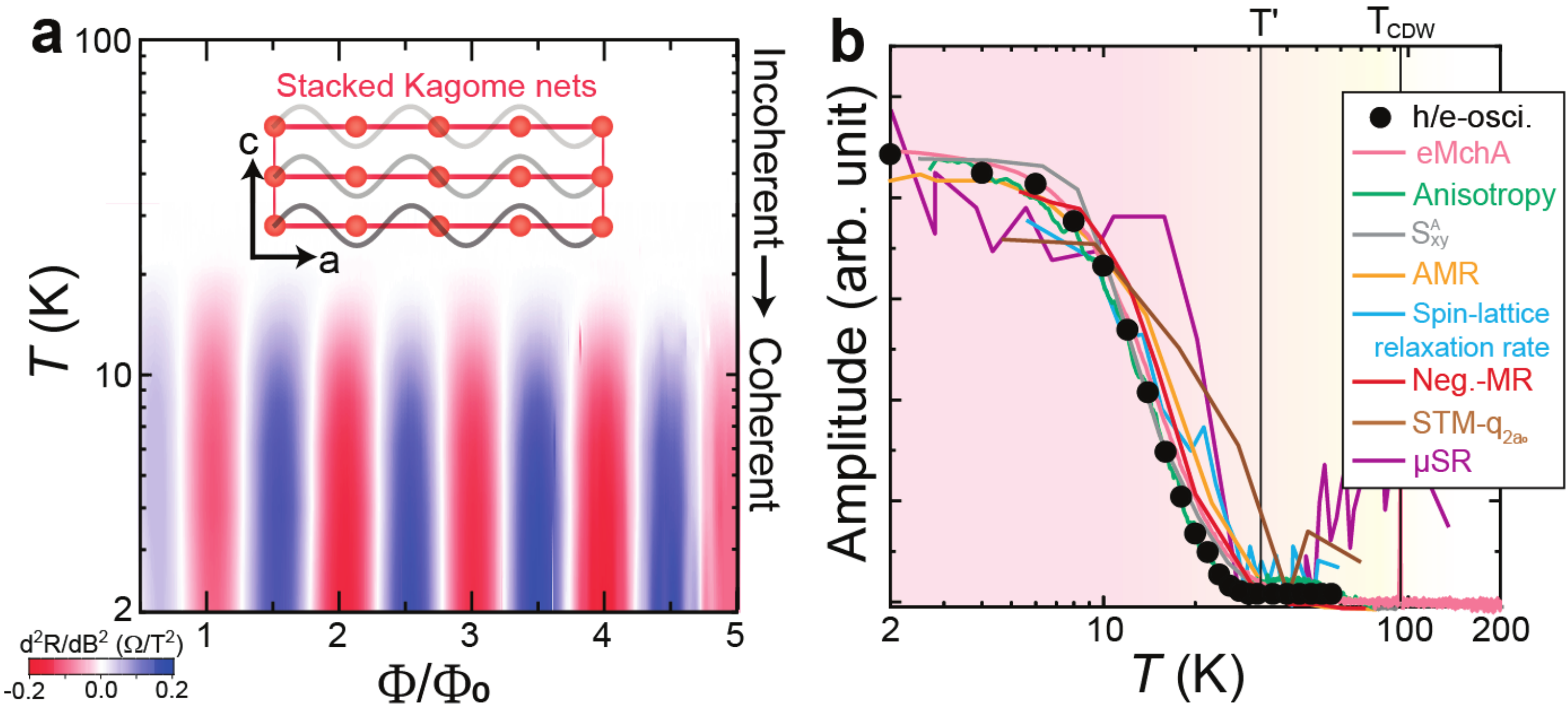}
		\caption{\textbf{Summary on temperature dependence and comparison with previous reports.} (a) $T$-dependence of $h/e$ oscillations in CsV$_3$Sb$_5$ in device S2. The amplitude becomes prominent below 20K, suggesting a crossover from quantum incoherent to coherent transport regime. (b) Summary of the previously reported temperature dependence of various results, including electric magneto-chiral anisotropy\cite{Guo2022}, field-induced in-plane conductivity anisotropy\cite{guo2024distinct}, anomalous Nernst effect\cite{DresdenThermal}, angular dependence of magnetoresistance\cite{xiang2021twofold}, negative magnetoresistance\cite{wei2024three}, spin-lattice relaxation rate\cite{Nie2022}, wavevector intensity of the 2$a_0$ charge order via STM measurements\cite{zhao2021cascade}, muon spin relaxation rate\cite{yu2021evidence}. All results fall on a universal temperature scaling below $T'\approx$ 30 K.}
	\label{Sum}
\end{figure}

The $h/e$ oscillations are remarkably stable against temperature and can clearly be resolved even at temperatures above 20~K despite the strongly reduced mean free path (Fig.~\ref{Sum}). Quantum oscillations are long washed out thermally, hence the vanishingly short Dingle mean free path can only be estimated by extrapolating the exponential tail of the Lifshitz-Kosevich behavior. Similarly, the most optimistic estimates of the transport mean free path from in-plane resistivity measurements reach only 150~nm at 20~K (see supplement). Thus the oscillations can be detected even when the quantum and transport mean free paths are substantially smaller than the device size, a challenge to theories rooted in semi-classics as well as single-particle coherence alike.

These unexpected findings remain at present theoretically unexplained, yet they are of natural importance for our understanding of CsV$_3$Sb$_5$, Kagome materials more generally, and quantum coherence in solids at large. 
While the appearance of an apparent coherence limited by sample size is a spectacular phenomenon in and of its own, the present experiment fundamentally prompts a question that may be phrased in the form of a paradox: how can the absence of well-resolved quantum oscillations be reconciled with the h/e oscillations in magnetoconductance given the fact that coherence lengths extracted in conventional ways for both phenomena are hugely incompatible?

Indeed, a hint toward an explanation stems from the temperature dependence (Fig.~\ref{Sum}). The amplitudes remain unaffected until about 10~K, above which they decay into the noise floor around 25~K. This temperature dependence quantitatively agrees with that of multiple electronic probes (Fig.~\ref{Sum}) including STM\cite{zhao2021cascade}, $\mu$SR\cite{yu2021evidence}, NMR\cite{Nie2022}, anomalous Nernst effect\cite{DresdenThermal}, electrical transport\cite{guo2024correlated,xiang2021twofold,wei2024three} and magnetochiral anisotropy\cite{Guo2022}. 
Further, the strong influence of the out-of-plane field component quantitatively mirrors field-switchable diode effects measured at much higher values of the in-plane field (see supplement) and chimes in the choir of experimental evidence showing field-switchability by out-of-plane fields, including STM\cite{Kchiral}, non-reciprocal transport\cite{Guo2022}, electronic anisotropy\cite{guo2024correlated} and magneto-optical Kerr effect\cite{liang2021three}. This remarkable agreement strongly suggests a substantial change in the electronic spectrum, which appears to support quantum coherence on the micron scale. 

\subsection*{Discussion and Outlook}

The h/e oscillations in the multiband system CsV$_3$Sb$_5$ appear within a charge ordered phase in contrast to the previously discovered PdCoO$_2$. This phase has been reported to have a high level of tunability — also by external magnetic fields —and is not fully characterized. In particular, no consensus is reached how precisely the Fermi surface reconstructs at low temperature due to the charge order. Even within these uncertainties, it is highly implausible that a non-interacting band structure can account for the observed phenomenology, as we demonstrated with our semiclassical analysis in terms of Bloch-Lorentz oscillations.

We are thus prompted to consider collective phenomena to provide a resource for the observed long-range coherence, especially by their onset at T'. These could be found in complex charge order akin to a loop or flux order, excitonic orders proposed to explain other aspects of CsV$_3$Sb$_5$\cite{scammell2023chiral,ingham2025vestigial}, a precursor state of superconductivity, or charge-transport contributions by a sliding density order. In any of these scenarios, the observed $h/e$ oscillations are supported by degrees of freedom distinct from the Landau quasiparticles contributing to quantum oscillations. Such a decoupling of degrees of freedom may be facilitated by the multiorbital nature of the material, where Sb $p$-electrons and V $d$-electrons reside on different bands and have been shown to participate distinctly in collective phenomena such as superconductivity\cite{hossain2024unconventional}. As a more quantitative exploration of such ideas did not lead to a satisfactory explanation of the data (see supplement), these observations are a case in point for a hitherto unknown source of remarkable long-range electron coherence.

Intriguingly, Kagome physics may prove to be a hidden key in the problem. In spite of the diametrically opposed physical situations in PdCoO$_2$ and \CVS, Kagome physics is a unifying theme. The Kagome structural motif is prone to enhanced correlations due to its orbital frustration on the three orbitals on the corner-sharing triangles. PdCoO$_2$ has been demonstrated to host a hidden Kagome structure, which, despite its hexagonal planar structure, arises from a three-fold orbital degeneracy that evolves around the Fermi surface\cite{usui2019hidden}. Its orbital degeneracy is intimately related with the three independent $h/e$ processes hosted in PdCoO$_2$~\cite{putzke2020h}. This notable genetic similarity tempts one into a speculative hypothesis, namely that frustration and topology in Kagome materials may provide a generic feature protecting the coherence between itinerant carriers, a conceptual and technological frontier alike.

\clearpage
\noindent \textbf{Acknowledgements: } We thank Julian P. Ingham, Ronny Thomale, Harley Scammell, Rafael M. Fernandes, Ady Stern, and Felix von Oppen for the insightful discussions. This work was funded by the European Research Council (ERC) under grant Free-Kagome  (Grant Agreement No. 101164280). This work was in part supported by the Deutsche Forschungsgemeinschaft under grants SFB 1143 (project-id 247310070) and the cluster of excellence ct.qmat (EXC 2147, project-id 390858490). This work was supported by HFML-RU/NWO-I, a member of the European Magnetic Field Laboratory (EMFL). A portion of this work was performed at the National High Magnetic Field Laboratory, which is supported by National Science Foundation Cooperative Agreement No. DMR-2128556*, the State of Florida, and the U.S. Department of Energy.

\noindent \textbf{Author Contributions:} Crystals were synthesized and characterized by D.C. and C.F.. The experiment design, FIB microstructuring, and magnetotransport measurements were performed by C.G., L.Z., C.P., and P.J.W.M.. M.R.v.D and S.W. assisted with the high-field magnetotransport measurements at the 35~T static magnet, while F.F.B. and R.M. helped with the electrical measurements in the pulsed magnet up to 65 T. K.W. performed the theoretical simulation based on the semiclassical Bloch-Lorentz model, and the DFT calculations were performed by M.G-A., M.A., I.E., and M.G.V.. The experimental results were analyzed by C.G. and P.J.W.M. All authors contributed to the writing of the paper.

\noindent \textbf{Competing Interests:} The authors declare that they have no competing financial interests.\\

\clearpage

\clearpage
\clearpage
\clearpage
\clearpage

%
\clearpage

\noindent \textbf{Data Availability:} Data supporting the findings of this study will be deposited to Zenodo with the access link displayed here.

\noindent \textbf{Code Availability:} Code that supports the findings of this study will be deposited to Zenodo with the access link displayed here.


\section*{References}
\begin{thebibliography}{52}%
	\makeatletter
	\providecommand \@ifxundefined [1]{%
		\@ifx{#1\undefined}
	}%
	\providecommand \@ifnum [1]{%
		\ifnum #1\expandafter \@firstoftwo
		\else \expandafter \@secondoftwo
		\fi
	}%
	\providecommand \@ifx [1]{%
		\ifx #1\expandafter \@firstoftwo
		\else \expandafter \@secondoftwo
		\fi
	}%
	\providecommand \natexlab [1]{#1}%
	\providecommand \enquote  [1]{``#1''}%
	\providecommand \bibnamefont  [1]{#1}%
	\providecommand \bibfnamefont [1]{#1}%
	\providecommand \citenamefont [1]{#1}%
	\providecommand \href@noop [0]{\@secondoftwo}%
	\providecommand \href [0]{\begingroup \@sanitize@url \@href}%
	\providecommand \@href[1]{\@@startlink{#1}\@@href}%
	\providecommand \@@href[1]{\endgroup#1\@@endlink}%
	\providecommand \@sanitize@url [0]{\catcode `\\12\catcode `\$12\catcode
		`\&12\catcode `\#12\catcode `\^12\catcode `\_12\catcode `\%12\relax}%
	\providecommand \@@startlink[1]{}%
	\providecommand \@@endlink[0]{}%
	\providecommand \url  [0]{\begingroup\@sanitize@url \@url }%
	\providecommand \@url [1]{\endgroup\@href {#1}{\urlprefix }}%
	\providecommand \urlprefix  [0]{URL }%
	\providecommand \Eprint [0]{\href }%
	\providecommand \doibase [0]{http://dx.doi.org/}%
	\providecommand \selectlanguage [0]{\@gobble}%
	\providecommand \bibinfo  [0]{\@secondoftwo}%
	\providecommand \bibfield  [0]{\@secondoftwo}%
	\providecommand \translation [1]{[#1]}%
	\providecommand \BibitemOpen [0]{}%
	\providecommand \bibitemStop [0]{}%
	\providecommand \bibitemNoStop [0]{.\EOS\space}%
	\providecommand \EOS [0]{\spacefactor3000\relax}%
	\providecommand \BibitemShut  [1]{\csname bibitem#1\endcsname}%
	\let\auto@bib@innerbib\@empty
	\bibitem [{\citenamefont {Davisson}\ and\ \citenamefont
		{Germer}(1927)}]{davisson1927scattering}%
	\BibitemOpen
	\bibfield  {author} {\bibinfo {author} {\bibfnamefont {C.}~\bibnamefont
			{Davisson}}\ and\ \bibinfo {author} {\bibfnamefont {L.~H.}\ \bibnamefont
			{Germer}},\ }\href@noop {} {\bibfield  {journal} {\bibinfo  {journal}
			{Nature}\ }\textbf {\bibinfo {volume} {119}},\ \bibinfo {pages} {558}
		(\bibinfo {year} {1927})}\BibitemShut {NoStop}%
	\bibitem [{\citenamefont {Olariu}\ and\ \citenamefont
		{Popescu}(1985)}]{olariu1985quantum}%
	\BibitemOpen
	\bibfield  {author} {\bibinfo {author} {\bibfnamefont {S.}~\bibnamefont
			{Olariu}}\ and\ \bibinfo {author} {\bibfnamefont {I.~I.}\ \bibnamefont
			{Popescu}},\ }\href@noop {} {\bibfield  {journal} {\bibinfo  {journal}
			{Reviews of Modern Physics}\ }\textbf {\bibinfo {volume} {57}},\ \bibinfo
		{pages} {339} (\bibinfo {year} {1985})}\BibitemShut {NoStop}%
	\bibitem [{\citenamefont {Washburn}\ and\ \citenamefont
		{Webb}(1986)}]{washburn1986aharonov}%
	\BibitemOpen
	\bibfield  {author} {\bibinfo {author} {\bibfnamefont {S.}~\bibnamefont
			{Washburn}}\ and\ \bibinfo {author} {\bibfnamefont {R.~A.}\ \bibnamefont
			{Webb}},\ }\href@noop {} {\bibfield  {journal} {\bibinfo  {journal} {Advances
				in Physics}\ }\textbf {\bibinfo {volume} {35}},\ \bibinfo {pages} {375}
		(\bibinfo {year} {1986})}\BibitemShut {NoStop}%
	\bibitem [{\citenamefont {Putzke}\ \emph {et~al.}(2020)\citenamefont {Putzke},
		\citenamefont {Bachmann}, \citenamefont {McGuinness}, \citenamefont
		{Zhakina}, \citenamefont {Sunko}, \citenamefont {Konczykowski}, \citenamefont
		{Oka}, \citenamefont {Moessner}, \citenamefont {Stern}, \citenamefont
		{K{\"o}nig} \emph {et~al.}}]{putzke2020h}%
	\BibitemOpen
	\bibfield  {author} {\bibinfo {author} {\bibfnamefont {C.}~\bibnamefont
			{Putzke}}, \bibinfo {author} {\bibfnamefont {M.~D.}\ \bibnamefont
			{Bachmann}}, \bibinfo {author} {\bibfnamefont {P.}~\bibnamefont
			{McGuinness}}, \bibinfo {author} {\bibfnamefont {E.}~\bibnamefont {Zhakina}},
		\bibinfo {author} {\bibfnamefont {V.}~\bibnamefont {Sunko}}, \bibinfo
		{author} {\bibfnamefont {M.}~\bibnamefont {Konczykowski}}, \bibinfo {author}
		{\bibfnamefont {T.}~\bibnamefont {Oka}}, \bibinfo {author} {\bibfnamefont
			{R.}~\bibnamefont {Moessner}}, \bibinfo {author} {\bibfnamefont
			{A.}~\bibnamefont {Stern}}, \bibinfo {author} {\bibfnamefont
			{M.}~\bibnamefont {K{\"o}nig}},  \emph {et~al.},\ }\href@noop {} {\bibfield
		{journal} {\bibinfo  {journal} {Science}\ }\textbf {\bibinfo {volume}
			{368}},\ \bibinfo {pages} {1234} (\bibinfo {year} {2020})}\BibitemShut
	{NoStop}%
	\bibitem [{\citenamefont {Bachmann}\ \emph {et~al.}(2022)\citenamefont
		{Bachmann}, \citenamefont {Sharpe}, \citenamefont {Baker}, \citenamefont
		{Barnard}, \citenamefont {Putzke}, \citenamefont {Scaffidi}, \citenamefont
		{Nandi}, \citenamefont {McGuinness}, \citenamefont {Zhakina}, \citenamefont
		{Moravec} \emph {et~al.}}]{bachmann2022directional}%
	\BibitemOpen
	\bibfield  {author} {\bibinfo {author} {\bibfnamefont {M.~D.}\ \bibnamefont
			{Bachmann}}, \bibinfo {author} {\bibfnamefont {A.~L.}\ \bibnamefont
			{Sharpe}}, \bibinfo {author} {\bibfnamefont {G.}~\bibnamefont {Baker}},
		\bibinfo {author} {\bibfnamefont {A.~W.}\ \bibnamefont {Barnard}}, \bibinfo
		{author} {\bibfnamefont {C.}~\bibnamefont {Putzke}}, \bibinfo {author}
		{\bibfnamefont {T.}~\bibnamefont {Scaffidi}}, \bibinfo {author}
		{\bibfnamefont {N.}~\bibnamefont {Nandi}}, \bibinfo {author} {\bibfnamefont
			{P.~H.}\ \bibnamefont {McGuinness}}, \bibinfo {author} {\bibfnamefont
			{E.}~\bibnamefont {Zhakina}}, \bibinfo {author} {\bibfnamefont
			{M.}~\bibnamefont {Moravec}},  \emph {et~al.},\ }\href@noop {} {\bibfield
		{journal} {\bibinfo  {journal} {Nature Physics}\ }\textbf {\bibinfo {volume}
			{18}},\ \bibinfo {pages} {819} (\bibinfo {year} {2022})}\BibitemShut
	{NoStop}%
	\bibitem [{\citenamefont {Bachmann}\ \emph
		{et~al.}(2019{\natexlab{a}})\citenamefont {Bachmann}, \citenamefont {Sharpe},
		\citenamefont {Barnard}, \citenamefont {Putzke}, \citenamefont {K{\"o}nig},
		\citenamefont {Khim}, \citenamefont {Goldhaber-Gordon}, \citenamefont
		{Mackenzie},\ and\ \citenamefont {Moll}}]{bachmann2019super}%
	\BibitemOpen
	\bibfield  {author} {\bibinfo {author} {\bibfnamefont {M.~D.}\ \bibnamefont
			{Bachmann}}, \bibinfo {author} {\bibfnamefont {A.~L.}\ \bibnamefont
			{Sharpe}}, \bibinfo {author} {\bibfnamefont {A.~W.}\ \bibnamefont {Barnard}},
		\bibinfo {author} {\bibfnamefont {C.}~\bibnamefont {Putzke}}, \bibinfo
		{author} {\bibfnamefont {M.}~\bibnamefont {K{\"o}nig}}, \bibinfo {author}
		{\bibfnamefont {S.}~\bibnamefont {Khim}}, \bibinfo {author} {\bibfnamefont
			{D.}~\bibnamefont {Goldhaber-Gordon}}, \bibinfo {author} {\bibfnamefont
			{A.~P.}\ \bibnamefont {Mackenzie}}, \ and\ \bibinfo {author} {\bibfnamefont
			{P.~J.}\ \bibnamefont {Moll}},\ }\href@noop {} {\bibfield  {journal}
		{\bibinfo  {journal} {Nature communications}\ }\textbf {\bibinfo {volume}
			{10}},\ \bibinfo {pages} {5081} (\bibinfo {year}
		{2019}{\natexlab{a}})}\BibitemShut {NoStop}%
	\bibitem [{\citenamefont {Nandi}\ \emph {et~al.}(2018)\citenamefont {Nandi},
		\citenamefont {Scaffidi}, \citenamefont {Kushwaha}, \citenamefont {Khim},
		\citenamefont {Barber}, \citenamefont {Sunko}, \citenamefont {Mazzola},
		\citenamefont {King}, \citenamefont {Rosner}, \citenamefont {Moll} \emph
		{et~al.}}]{nandi2018unconventional}%
	\BibitemOpen
	\bibfield  {author} {\bibinfo {author} {\bibfnamefont {N.}~\bibnamefont
			{Nandi}}, \bibinfo {author} {\bibfnamefont {T.}~\bibnamefont {Scaffidi}},
		\bibinfo {author} {\bibfnamefont {P.}~\bibnamefont {Kushwaha}}, \bibinfo
		{author} {\bibfnamefont {S.}~\bibnamefont {Khim}}, \bibinfo {author}
		{\bibfnamefont {M.~E.}\ \bibnamefont {Barber}}, \bibinfo {author}
		{\bibfnamefont {V.}~\bibnamefont {Sunko}}, \bibinfo {author} {\bibfnamefont
			{F.}~\bibnamefont {Mazzola}}, \bibinfo {author} {\bibfnamefont {P.~D.}\
			\bibnamefont {King}}, \bibinfo {author} {\bibfnamefont {H.}~\bibnamefont
			{Rosner}}, \bibinfo {author} {\bibfnamefont {P.~J.}\ \bibnamefont {Moll}},
		\emph {et~al.},\ }\href@noop {} {\bibfield  {journal} {\bibinfo  {journal}
			{npj Quantum Materials}\ }\textbf {\bibinfo {volume} {3}},\ \bibinfo {pages}
		{66} (\bibinfo {year} {2018})}\BibitemShut {NoStop}%
	\bibitem [{\citenamefont {Moll}\ \emph {et~al.}(2016)\citenamefont {Moll},
		\citenamefont {Kushwaha}, \citenamefont {Nandi}, \citenamefont {Schmidt},\
		and\ \citenamefont {Mackenzie}}]{moll2016evidence}%
	\BibitemOpen
	\bibfield  {author} {\bibinfo {author} {\bibfnamefont {P.~J.}\ \bibnamefont
			{Moll}}, \bibinfo {author} {\bibfnamefont {P.}~\bibnamefont {Kushwaha}},
		\bibinfo {author} {\bibfnamefont {N.}~\bibnamefont {Nandi}}, \bibinfo
		{author} {\bibfnamefont {B.}~\bibnamefont {Schmidt}}, \ and\ \bibinfo
		{author} {\bibfnamefont {A.~P.}\ \bibnamefont {Mackenzie}},\ }\href@noop {}
	{\bibfield  {journal} {\bibinfo  {journal} {Science}\ }\textbf {\bibinfo
			{volume} {351}},\ \bibinfo {pages} {1061} (\bibinfo {year}
		{2016})}\BibitemShut {NoStop}%
	\bibitem [{\citenamefont {Ortiz}\ \emph {et~al.}(2019)\citenamefont {Ortiz},
		\citenamefont {Gomes}, \citenamefont {Morey}, \citenamefont {Winiarski},
		\citenamefont {Bordelon}, \citenamefont {Mangum}, \citenamefont {Oswald},
		\citenamefont {Rodriguez-Rivera}, \citenamefont {Neilson}, \citenamefont
		{Wilson}, \citenamefont {Ertekin}, \citenamefont {McQueen},\ and\
		\citenamefont {Toberer}}]{ortiz2019new}%
	\BibitemOpen
	\bibfield  {author} {\bibinfo {author} {\bibfnamefont {B.~R.}\ \bibnamefont
			{Ortiz}}, \bibinfo {author} {\bibfnamefont {L.~C.}\ \bibnamefont {Gomes}},
		\bibinfo {author} {\bibfnamefont {J.~R.}\ \bibnamefont {Morey}}, \bibinfo
		{author} {\bibfnamefont {M.}~\bibnamefont {Winiarski}}, \bibinfo {author}
		{\bibfnamefont {M.}~\bibnamefont {Bordelon}}, \bibinfo {author}
		{\bibfnamefont {J.~S.}\ \bibnamefont {Mangum}}, \bibinfo {author}
		{\bibfnamefont {I.~W.}\ \bibnamefont {Oswald}}, \bibinfo {author}
		{\bibfnamefont {J.~A.}\ \bibnamefont {Rodriguez-Rivera}}, \bibinfo {author}
		{\bibfnamefont {J.~R.}\ \bibnamefont {Neilson}}, \bibinfo {author}
		{\bibfnamefont {S.~D.}\ \bibnamefont {Wilson}}, \bibinfo {author}
		{\bibfnamefont {E.}~\bibnamefont {Ertekin}}, \bibinfo {author} {\bibfnamefont
			{T.~M.}\ \bibnamefont {McQueen}}, \ and\ \bibinfo {author} {\bibfnamefont
			{E.~S.}\ \bibnamefont {Toberer}},\ }\href@noop {} {\bibfield  {journal}
		{\bibinfo  {journal} {Phys. Rev. Mater.}\ }\textbf {\bibinfo {volume} {3}},\
		\bibinfo {pages} {094407} (\bibinfo {year} {2019})}\BibitemShut {NoStop}%
	\bibitem [{\citenamefont {Ortiz}\ \emph {et~al.}(2020)\citenamefont {Ortiz},
		\citenamefont {Teicher}, \citenamefont {Hu}, \citenamefont {Zuo},
		\citenamefont {Sarte}, \citenamefont {Schueller}, \citenamefont {Abeykoon},
		\citenamefont {Krogstad}, \citenamefont {Rosenkranz}, \citenamefont {Osborn},
		\citenamefont {Seshadri}, \citenamefont {Balents}, \citenamefont {He},\ and\
		\citenamefont {Wilson}}]{ortiz2020cs}%
	\BibitemOpen
	\bibfield  {author} {\bibinfo {author} {\bibfnamefont {B.~R.}\ \bibnamefont
			{Ortiz}}, \bibinfo {author} {\bibfnamefont {S.~M.}\ \bibnamefont {Teicher}},
		\bibinfo {author} {\bibfnamefont {Y.}~\bibnamefont {Hu}}, \bibinfo {author}
		{\bibfnamefont {J.~L.}\ \bibnamefont {Zuo}}, \bibinfo {author} {\bibfnamefont
			{P.~M.}\ \bibnamefont {Sarte}}, \bibinfo {author} {\bibfnamefont {E.~C.}\
			\bibnamefont {Schueller}}, \bibinfo {author} {\bibfnamefont {A.~M.}\
			\bibnamefont {Abeykoon}}, \bibinfo {author} {\bibfnamefont {M.~J.}\
			\bibnamefont {Krogstad}}, \bibinfo {author} {\bibfnamefont {S.}~\bibnamefont
			{Rosenkranz}}, \bibinfo {author} {\bibfnamefont {R.}~\bibnamefont {Osborn}},
		\bibinfo {author} {\bibfnamefont {R.}~\bibnamefont {Seshadri}}, \bibinfo
		{author} {\bibfnamefont {L.}~\bibnamefont {Balents}}, \bibinfo {author}
		{\bibfnamefont {J.}~\bibnamefont {He}}, \ and\ \bibinfo {author}
		{\bibfnamefont {S.~D.}\ \bibnamefont {Wilson}},\ }\href@noop {} {\bibfield
		{journal} {\bibinfo  {journal} {Phys. Rev. Lett.}\ }\textbf {\bibinfo
			{volume} {125}},\ \bibinfo {pages} {247002} (\bibinfo {year}
		{2020})}\BibitemShut {NoStop}%
	\bibitem [{\citenamefont {Xie}\ \emph {et~al.}(2022{\natexlab{a}})\citenamefont
		{Xie}, \citenamefont {Li}, \citenamefont {Bourges}, \citenamefont {Ivanov},
		\citenamefont {Ye}, \citenamefont {Yin}, \citenamefont {Hasan}, \citenamefont
		{Luo}, \citenamefont {Yao}, \citenamefont {Wang}, \citenamefont {Xu},\ and\
		\citenamefont {Dai}}]{epc}%
	\BibitemOpen
	\bibfield  {author} {\bibinfo {author} {\bibfnamefont {Y.}~\bibnamefont
			{Xie}}, \bibinfo {author} {\bibfnamefont {Y.}~\bibnamefont {Li}}, \bibinfo
		{author} {\bibfnamefont {P.}~\bibnamefont {Bourges}}, \bibinfo {author}
		{\bibfnamefont {A.}~\bibnamefont {Ivanov}}, \bibinfo {author} {\bibfnamefont
			{Z.}~\bibnamefont {Ye}}, \bibinfo {author} {\bibfnamefont {J.-X.}\
			\bibnamefont {Yin}}, \bibinfo {author} {\bibfnamefont {M.~Z.}\ \bibnamefont
			{Hasan}}, \bibinfo {author} {\bibfnamefont {A.}~\bibnamefont {Luo}}, \bibinfo
		{author} {\bibfnamefont {Y.}~\bibnamefont {Yao}}, \bibinfo {author}
		{\bibfnamefont {Z.}~\bibnamefont {Wang}}, \bibinfo {author} {\bibfnamefont
			{G.}~\bibnamefont {Xu}}, \ and\ \bibinfo {author} {\bibfnamefont
			{P.}~\bibnamefont {Dai}},\ }\href@noop {} {\bibfield  {journal} {\bibinfo
			{journal} {Phys. Rev. B}\ }\textbf {\bibinfo {volume} {105}},\ \bibinfo
		{pages} {L140501} (\bibinfo {year} {2022}{\natexlab{a}})}\BibitemShut
	{NoStop}%
	\bibitem [{\citenamefont {Gutierrez-Amigo}\ \emph {et~al.}(2024)\citenamefont
		{Gutierrez-Amigo}, \citenamefont {Dangi{\'c}}, \citenamefont {Guo},
		\citenamefont {Felser}, \citenamefont {Moll}, \citenamefont {Vergniory},\
		and\ \citenamefont {Errea}}]{gutierrez2024phonon}%
	\BibitemOpen
	\bibfield  {author} {\bibinfo {author} {\bibfnamefont {M.}~\bibnamefont
			{Gutierrez-Amigo}}, \bibinfo {author} {\bibfnamefont {D.}~\bibnamefont
			{Dangi{\'c}}}, \bibinfo {author} {\bibfnamefont {C.}~\bibnamefont {Guo}},
		\bibinfo {author} {\bibfnamefont {C.}~\bibnamefont {Felser}}, \bibinfo
		{author} {\bibfnamefont {P.~J.}\ \bibnamefont {Moll}}, \bibinfo {author}
		{\bibfnamefont {M.~G.}\ \bibnamefont {Vergniory}}, \ and\ \bibinfo {author}
		{\bibfnamefont {I.}~\bibnamefont {Errea}},\ }\href@noop {} {\bibfield
		{journal} {\bibinfo  {journal} {Communications Materials}\ }\textbf {\bibinfo
			{volume} {5}},\ \bibinfo {pages} {234} (\bibinfo {year} {2024})}\BibitemShut
	{NoStop}%
	\bibitem [{\citenamefont {Han}\ \emph {et~al.}(2023)\citenamefont {Han},
		\citenamefont {Che}, \citenamefont {Ye},\ and\ \citenamefont
		{Huang}}]{Sbcris}%
	\BibitemOpen
	\bibfield  {author} {\bibinfo {author} {\bibfnamefont {T.}~\bibnamefont
			{Han}}, \bibinfo {author} {\bibfnamefont {J.}~\bibnamefont {Che}}, \bibinfo
		{author} {\bibfnamefont {C.}~\bibnamefont {Ye}}, \ and\ \bibinfo {author}
		{\bibfnamefont {H.}~\bibnamefont {Huang}},\ }\href@noop {} {\bibfield
		{journal} {\bibinfo  {journal} {Crystals}\ }\textbf {\bibinfo {volume}
			{13}},\ \bibinfo {pages} {321} (\bibinfo {year} {2023})}\BibitemShut
	{NoStop}%
	\bibitem [{\citenamefont {Gupta}\ \emph {et~al.}(2022)\citenamefont {Gupta},
		\citenamefont {Das}, \citenamefont {Mielke~III}, \citenamefont {Guguchia},
		\citenamefont {Shiroka}, \citenamefont {Baines}, \citenamefont {Bartkowiak},
		\citenamefont {Luetkens}, \citenamefont {Khasanov}, \citenamefont {Yin} \emph
		{et~al.}}]{gupta2022microscopic}%
	\BibitemOpen
	\bibfield  {author} {\bibinfo {author} {\bibfnamefont {R.}~\bibnamefont
			{Gupta}}, \bibinfo {author} {\bibfnamefont {D.}~\bibnamefont {Das}}, \bibinfo
		{author} {\bibfnamefont {C.~H.}\ \bibnamefont {Mielke~III}}, \bibinfo
		{author} {\bibfnamefont {Z.}~\bibnamefont {Guguchia}}, \bibinfo {author}
		{\bibfnamefont {T.}~\bibnamefont {Shiroka}}, \bibinfo {author} {\bibfnamefont
			{C.}~\bibnamefont {Baines}}, \bibinfo {author} {\bibfnamefont
			{M.}~\bibnamefont {Bartkowiak}}, \bibinfo {author} {\bibfnamefont
			{H.}~\bibnamefont {Luetkens}}, \bibinfo {author} {\bibfnamefont
			{R.}~\bibnamefont {Khasanov}}, \bibinfo {author} {\bibfnamefont
			{Q.}~\bibnamefont {Yin}},  \emph {et~al.},\ }\href@noop {} {\bibfield
		{journal} {\bibinfo  {journal} {npj Quantum Materials}\ }\textbf {\bibinfo
			{volume} {7}},\ \bibinfo {pages} {49} (\bibinfo {year} {2022})}\BibitemShut
	{NoStop}%
	\bibitem [{\citenamefont {Liang}\ \emph {et~al.}(2021)\citenamefont {Liang},
		\citenamefont {Hou}, \citenamefont {Zhang}, \citenamefont {Ma}, \citenamefont
		{Wu}, \citenamefont {Zhang}, \citenamefont {Yu}, \citenamefont {Ying},
		\citenamefont {Jiang}, \citenamefont {Shan}, \citenamefont {Wang},\ and\
		\citenamefont {Chen}}]{liang2021three}%
	\BibitemOpen
	\bibfield  {author} {\bibinfo {author} {\bibfnamefont {Z.}~\bibnamefont
			{Liang}}, \bibinfo {author} {\bibfnamefont {X.}~\bibnamefont {Hou}}, \bibinfo
		{author} {\bibfnamefont {F.}~\bibnamefont {Zhang}}, \bibinfo {author}
		{\bibfnamefont {W.}~\bibnamefont {Ma}}, \bibinfo {author} {\bibfnamefont
			{P.}~\bibnamefont {Wu}}, \bibinfo {author} {\bibfnamefont {Z.}~\bibnamefont
			{Zhang}}, \bibinfo {author} {\bibfnamefont {F.}~\bibnamefont {Yu}}, \bibinfo
		{author} {\bibfnamefont {J.-J.}\ \bibnamefont {Ying}}, \bibinfo {author}
		{\bibfnamefont {K.}~\bibnamefont {Jiang}}, \bibinfo {author} {\bibfnamefont
			{L.}~\bibnamefont {Shan}}, \bibinfo {author} {\bibfnamefont {Z.}~\bibnamefont
			{Wang}}, \ and\ \bibinfo {author} {\bibfnamefont {X.}~\bibnamefont {Chen}},\
	}\href@noop {} {\bibfield  {journal} {\bibinfo  {journal} {Phys. Rev. X}\
		}\textbf {\bibinfo {volume} {11}},\ \bibinfo {pages} {031026} (\bibinfo
		{year} {2021})}\BibitemShut {NoStop}%
	\bibitem [{\citenamefont {Zhang}\ \emph {et~al.}(2023)\citenamefont {Zhang},
		\citenamefont {Liu}, \citenamefont {Wang}, \citenamefont {Tsang},
		\citenamefont {Wang}, \citenamefont {Lam}, \citenamefont {Wang},
		\citenamefont {Xie}, \citenamefont {Zhou}, \citenamefont {Zhao} \emph
		{et~al.}}]{zhang2023nodeless}%
	\BibitemOpen
	\bibfield  {author} {\bibinfo {author} {\bibfnamefont {W.}~\bibnamefont
			{Zhang}}, \bibinfo {author} {\bibfnamefont {X.}~\bibnamefont {Liu}}, \bibinfo
		{author} {\bibfnamefont {L.}~\bibnamefont {Wang}}, \bibinfo {author}
		{\bibfnamefont {C.~W.}\ \bibnamefont {Tsang}}, \bibinfo {author}
		{\bibfnamefont {Z.}~\bibnamefont {Wang}}, \bibinfo {author} {\bibfnamefont
			{S.~T.}\ \bibnamefont {Lam}}, \bibinfo {author} {\bibfnamefont
			{W.}~\bibnamefont {Wang}}, \bibinfo {author} {\bibfnamefont {J.}~\bibnamefont
			{Xie}}, \bibinfo {author} {\bibfnamefont {X.}~\bibnamefont {Zhou}}, \bibinfo
		{author} {\bibfnamefont {Y.}~\bibnamefont {Zhao}},  \emph {et~al.},\
	}\href@noop {} {\bibfield  {journal} {\bibinfo  {journal} {Nano Letters}\
		}\textbf {\bibinfo {volume} {23}},\ \bibinfo {pages} {872} (\bibinfo {year}
		{2023})}\BibitemShut {NoStop}%
	\bibitem [{\citenamefont {Zhao}\ \emph {et~al.}(2024)\citenamefont {Zhao},
		\citenamefont {Wang}, \citenamefont {Xia}, \citenamefont {Yin}, \citenamefont
		{Deng}, \citenamefont {Liu}, \citenamefont {Liu}, \citenamefont {Zhang},
		\citenamefont {Ni}, \citenamefont {Huang} \emph {et~al.}}]{zhao2024ultralow}%
	\BibitemOpen
	\bibfield  {author} {\bibinfo {author} {\bibfnamefont {C.}~\bibnamefont
			{Zhao}}, \bibinfo {author} {\bibfnamefont {L.}~\bibnamefont {Wang}}, \bibinfo
		{author} {\bibfnamefont {W.}~\bibnamefont {Xia}}, \bibinfo {author}
		{\bibfnamefont {Q.}~\bibnamefont {Yin}}, \bibinfo {author} {\bibfnamefont
			{H.}~\bibnamefont {Deng}}, \bibinfo {author} {\bibfnamefont {G.}~\bibnamefont
			{Liu}}, \bibinfo {author} {\bibfnamefont {J.}~\bibnamefont {Liu}}, \bibinfo
		{author} {\bibfnamefont {X.}~\bibnamefont {Zhang}}, \bibinfo {author}
		{\bibfnamefont {J.}~\bibnamefont {Ni}}, \bibinfo {author} {\bibfnamefont
			{Y.}~\bibnamefont {Huang}},  \emph {et~al.},\ }\href@noop {} {\bibfield
		{journal} {\bibinfo  {journal} {Chinese Physics Letters}\ } (\bibinfo {year}
		{2024})}\BibitemShut {NoStop}%
	\bibitem [{\citenamefont {Le}\ \emph {et~al.}(2024)\citenamefont {Le},
		\citenamefont {Pan}, \citenamefont {Xu}, \citenamefont {Liu}, \citenamefont
		{Wang}, \citenamefont {Lou}, \citenamefont {Yang}, \citenamefont {Wang},
		\citenamefont {Yao}, \citenamefont {Wu} \emph
		{et~al.}}]{le2024superconducting}%
	\BibitemOpen
	\bibfield  {author} {\bibinfo {author} {\bibfnamefont {T.}~\bibnamefont
			{Le}}, \bibinfo {author} {\bibfnamefont {Z.}~\bibnamefont {Pan}}, \bibinfo
		{author} {\bibfnamefont {Z.}~\bibnamefont {Xu}}, \bibinfo {author}
		{\bibfnamefont {J.}~\bibnamefont {Liu}}, \bibinfo {author} {\bibfnamefont
			{J.}~\bibnamefont {Wang}}, \bibinfo {author} {\bibfnamefont {Z.}~\bibnamefont
			{Lou}}, \bibinfo {author} {\bibfnamefont {X.}~\bibnamefont {Yang}}, \bibinfo
		{author} {\bibfnamefont {Z.}~\bibnamefont {Wang}}, \bibinfo {author}
		{\bibfnamefont {Y.}~\bibnamefont {Yao}}, \bibinfo {author} {\bibfnamefont
			{C.}~\bibnamefont {Wu}},  \emph {et~al.},\ }\href@noop {} {\bibfield
		{journal} {\bibinfo  {journal} {Nature}\ ,\ \bibinfo {pages} {1}} (\bibinfo
		{year} {2024})}\BibitemShut {NoStop}%
	\bibitem [{\citenamefont {Hossain}\ \emph {et~al.}(2025)\citenamefont
		{Hossain}, \citenamefont {Zhang}, \citenamefont {Choi}, \citenamefont
		{Ratkovski}, \citenamefont {L{\"u}scher}, \citenamefont {Li}, \citenamefont
		{Jiang}, \citenamefont {Litskevich}, \citenamefont {Cheng}, \citenamefont
		{Yin} \emph {et~al.}}]{hossain2024unconventional}%
	\BibitemOpen
	\bibfield  {author} {\bibinfo {author} {\bibfnamefont {M.~S.}\ \bibnamefont
			{Hossain}}, \bibinfo {author} {\bibfnamefont {Q.}~\bibnamefont {Zhang}},
		\bibinfo {author} {\bibfnamefont {E.~S.}\ \bibnamefont {Choi}}, \bibinfo
		{author} {\bibfnamefont {D.}~\bibnamefont {Ratkovski}}, \bibinfo {author}
		{\bibfnamefont {B.}~\bibnamefont {L{\"u}scher}}, \bibinfo {author}
		{\bibfnamefont {Y.}~\bibnamefont {Li}}, \bibinfo {author} {\bibfnamefont
			{Y.-X.}\ \bibnamefont {Jiang}}, \bibinfo {author} {\bibfnamefont
			{M.}~\bibnamefont {Litskevich}}, \bibinfo {author} {\bibfnamefont {Z.-J.}\
			\bibnamefont {Cheng}}, \bibinfo {author} {\bibfnamefont {J.-X.}\ \bibnamefont
			{Yin}},  \emph {et~al.},\ }\href@noop {} {\bibfield  {journal} {\bibinfo
			{journal} {Nature Physics}\ ,\ \bibinfo {pages} {1}} (\bibinfo {year}
		{2025})}\BibitemShut {NoStop}%
	\bibitem [{\citenamefont {Guo}\ \emph {et~al.}(2022)\citenamefont {Guo},
		\citenamefont {Putzke}, \citenamefont {Konyzheva}, \citenamefont {Huang},
		\citenamefont {Gutierrez-Amigo}, \citenamefont {Errea}, \citenamefont {Chen},
		\citenamefont {Vergniory}, \citenamefont {Felser}, \citenamefont {Fischer},
		\citenamefont {Neupert},\ and\ \citenamefont {Moll}}]{Guo2022}%
	\BibitemOpen
	\bibfield  {author} {\bibinfo {author} {\bibfnamefont {C.}~\bibnamefont
			{Guo}}, \bibinfo {author} {\bibfnamefont {C.}~\bibnamefont {Putzke}},
		\bibinfo {author} {\bibfnamefont {S.}~\bibnamefont {Konyzheva}}, \bibinfo
		{author} {\bibfnamefont {X.}~\bibnamefont {Huang}}, \bibinfo {author}
		{\bibfnamefont {M.}~\bibnamefont {Gutierrez-Amigo}}, \bibinfo {author}
		{\bibfnamefont {I.}~\bibnamefont {Errea}}, \bibinfo {author} {\bibfnamefont
			{D.}~\bibnamefont {Chen}}, \bibinfo {author} {\bibfnamefont {M.~G.}\
			\bibnamefont {Vergniory}}, \bibinfo {author} {\bibfnamefont {C.}~\bibnamefont
			{Felser}}, \bibinfo {author} {\bibfnamefont {M.~H.}\ \bibnamefont {Fischer}},
		\bibinfo {author} {\bibfnamefont {T.}~\bibnamefont {Neupert}}, \ and\
		\bibinfo {author} {\bibfnamefont {P.~J.~W.}\ \bibnamefont {Moll}},\
	}\href@noop {} {\bibfield  {journal} {\bibinfo  {journal} {Nature}\ }\textbf
		{\bibinfo {volume} {611}},\ \bibinfo {pages} {461} (\bibinfo {year}
		{2022})}\BibitemShut {NoStop}%
	\bibitem [{\citenamefont {Xiang}\ \emph {et~al.}(2021)\citenamefont {Xiang},
		\citenamefont {Li}, \citenamefont {Li}, \citenamefont {Xie}, \citenamefont
		{Yang}, \citenamefont {Wang}, \citenamefont {Yao},\ and\ \citenamefont
		{Wen}}]{xiang2021twofold}%
	\BibitemOpen
	\bibfield  {author} {\bibinfo {author} {\bibfnamefont {Y.}~\bibnamefont
			{Xiang}}, \bibinfo {author} {\bibfnamefont {Q.}~\bibnamefont {Li}}, \bibinfo
		{author} {\bibfnamefont {Y.}~\bibnamefont {Li}}, \bibinfo {author}
		{\bibfnamefont {W.}~\bibnamefont {Xie}}, \bibinfo {author} {\bibfnamefont
			{H.}~\bibnamefont {Yang}}, \bibinfo {author} {\bibfnamefont {Z.}~\bibnamefont
			{Wang}}, \bibinfo {author} {\bibfnamefont {Y.}~\bibnamefont {Yao}}, \ and\
		\bibinfo {author} {\bibfnamefont {H.-H.}\ \bibnamefont {Wen}},\ }\href@noop
	{} {\bibfield  {journal} {\bibinfo  {journal} {Nat. Commun.}\ }\textbf
		{\bibinfo {volume} {12}},\ \bibinfo {pages} {6727} (\bibinfo {year}
		{2021})}\BibitemShut {NoStop}%
	\bibitem [{\citenamefont {Guo}\ \emph {et~al.}(2024{\natexlab{a}})\citenamefont
		{Guo}, \citenamefont {Wagner}, \citenamefont {Putzke}, \citenamefont {Chen},
		\citenamefont {Wang}, \citenamefont {Zhang}, \citenamefont {Gutierrez-Amigo},
		\citenamefont {Errea}, \citenamefont {Vergniory}, \citenamefont {Felser}
		\emph {et~al.}}]{guo2024correlated}%
	\BibitemOpen
	\bibfield  {author} {\bibinfo {author} {\bibfnamefont {C.}~\bibnamefont
			{Guo}}, \bibinfo {author} {\bibfnamefont {G.}~\bibnamefont {Wagner}},
		\bibinfo {author} {\bibfnamefont {C.}~\bibnamefont {Putzke}}, \bibinfo
		{author} {\bibfnamefont {D.}~\bibnamefont {Chen}}, \bibinfo {author}
		{\bibfnamefont {K.}~\bibnamefont {Wang}}, \bibinfo {author} {\bibfnamefont
			{L.}~\bibnamefont {Zhang}}, \bibinfo {author} {\bibfnamefont
			{M.}~\bibnamefont {Gutierrez-Amigo}}, \bibinfo {author} {\bibfnamefont
			{I.}~\bibnamefont {Errea}}, \bibinfo {author} {\bibfnamefont {M.~G.}\
			\bibnamefont {Vergniory}}, \bibinfo {author} {\bibfnamefont {C.}~\bibnamefont
			{Felser}},  \emph {et~al.},\ }\href@noop {} {\bibfield  {journal} {\bibinfo
			{journal} {Nature Physics}\ }\textbf {\bibinfo {volume} {20}},\ \bibinfo
		{pages} {579} (\bibinfo {year} {2024}{\natexlab{a}})}\BibitemShut {NoStop}%
	\bibitem [{\citenamefont {Chen}\ \emph {et~al.}(2021)\citenamefont {Chen},
		\citenamefont {He}, \citenamefont {Yao}, \citenamefont {Pan}, \citenamefont
		{Lin}, \citenamefont {Schnelle}, \citenamefont {Sun}, \citenamefont {Gooth},
		\citenamefont {Taillefer},\ and\ \citenamefont {Felser}}]{DresdenThermal}%
	\BibitemOpen
	\bibfield  {author} {\bibinfo {author} {\bibfnamefont {D.}~\bibnamefont
			{Chen}}, \bibinfo {author} {\bibfnamefont {B.}~\bibnamefont {He}}, \bibinfo
		{author} {\bibfnamefont {M.}~\bibnamefont {Yao}}, \bibinfo {author}
		{\bibfnamefont {Y.}~\bibnamefont {Pan}}, \bibinfo {author} {\bibfnamefont
			{H.}~\bibnamefont {Lin}}, \bibinfo {author} {\bibfnamefont {W.}~\bibnamefont
			{Schnelle}}, \bibinfo {author} {\bibfnamefont {Y.}~\bibnamefont {Sun}},
		\bibinfo {author} {\bibfnamefont {J.}~\bibnamefont {Gooth}}, \bibinfo
		{author} {\bibfnamefont {L.}~\bibnamefont {Taillefer}}, \ and\ \bibinfo
		{author} {\bibfnamefont {C.}~\bibnamefont {Felser}},\ }\href@noop {}
	{\bibfield  {journal} {\bibinfo  {journal} {arXiv:2110.13085}\ } (\bibinfo
		{year} {2021})}\BibitemShut {NoStop}%
	\bibitem [{\citenamefont {Yu}\ \emph {et~al.}(2021)\citenamefont {Yu},
		\citenamefont {Wang}, \citenamefont {Zhang}, \citenamefont {Sander},
		\citenamefont {Ni}, \citenamefont {Lu}, \citenamefont {Ma}, \citenamefont
		{Wang}, \citenamefont {Zhao}, \citenamefont {Chen}, \citenamefont {Jiang},
		\citenamefont {Zhang}, \citenamefont {Yang}, \citenamefont {Zhou},
		\citenamefont {Dong}, \citenamefont {Johnson}, \citenamefont {Graf},
		\citenamefont {Hu}, \citenamefont {Gao},\ and\ \citenamefont
		{Zhao}}]{yu2021evidence}%
	\BibitemOpen
	\bibfield  {author} {\bibinfo {author} {\bibfnamefont {L.}~\bibnamefont
			{Yu}}, \bibinfo {author} {\bibfnamefont {C.}~\bibnamefont {Wang}}, \bibinfo
		{author} {\bibfnamefont {Y.}~\bibnamefont {Zhang}}, \bibinfo {author}
		{\bibfnamefont {M.}~\bibnamefont {Sander}}, \bibinfo {author} {\bibfnamefont
			{S.}~\bibnamefont {Ni}}, \bibinfo {author} {\bibfnamefont {Z.}~\bibnamefont
			{Lu}}, \bibinfo {author} {\bibfnamefont {S.}~\bibnamefont {Ma}}, \bibinfo
		{author} {\bibfnamefont {Z.}~\bibnamefont {Wang}}, \bibinfo {author}
		{\bibfnamefont {Z.}~\bibnamefont {Zhao}}, \bibinfo {author} {\bibfnamefont
			{H.}~\bibnamefont {Chen}}, \bibinfo {author} {\bibfnamefont {K.}~\bibnamefont
			{Jiang}}, \bibinfo {author} {\bibfnamefont {Y.}~\bibnamefont {Zhang}},
		\bibinfo {author} {\bibfnamefont {H.}~\bibnamefont {Yang}}, \bibinfo {author}
		{\bibfnamefont {F.}~\bibnamefont {Zhou}}, \bibinfo {author} {\bibfnamefont
			{X.}~\bibnamefont {Dong}}, \bibinfo {author} {\bibfnamefont {S.~L.}\
			\bibnamefont {Johnson}}, \bibinfo {author} {\bibfnamefont {M.~J.}\
			\bibnamefont {Graf}}, \bibinfo {author} {\bibfnamefont {J.}~\bibnamefont
			{Hu}}, \bibinfo {author} {\bibfnamefont {H.-J.}\ \bibnamefont {Gao}}, \ and\
		\bibinfo {author} {\bibfnamefont {Z.}~\bibnamefont {Zhao}},\ }\href@noop {}
	{\bibfield  {journal} {\bibinfo  {journal} {arXiv:2107.10714}\ } (\bibinfo
		{year} {2021})}\BibitemShut {NoStop}%
	\bibitem [{\citenamefont {Nie}\ \emph {et~al.}(2022)\citenamefont {Nie},
		\citenamefont {Sun}, \citenamefont {Ma}, \citenamefont {Song}, \citenamefont
		{Zheng}, \citenamefont {Liang}, \citenamefont {Wu}, \citenamefont {Yu},
		\citenamefont {Li}, \citenamefont {Shan}, \citenamefont {Zhao}, \citenamefont
		{Li}, \citenamefont {Kang}, \citenamefont {Wu}, \citenamefont {Zhou},
		\citenamefont {Liu}, \citenamefont {Xiang}, \citenamefont {Ying},
		\citenamefont {Wang}, \citenamefont {Wu},\ and\ \citenamefont
		{Chen}}]{Nie2022}%
	\BibitemOpen
	\bibfield  {author} {\bibinfo {author} {\bibfnamefont {L.}~\bibnamefont
			{Nie}}, \bibinfo {author} {\bibfnamefont {K.}~\bibnamefont {Sun}}, \bibinfo
		{author} {\bibfnamefont {W.}~\bibnamefont {Ma}}, \bibinfo {author}
		{\bibfnamefont {D.}~\bibnamefont {Song}}, \bibinfo {author} {\bibfnamefont
			{L.}~\bibnamefont {Zheng}}, \bibinfo {author} {\bibfnamefont
			{Z.}~\bibnamefont {Liang}}, \bibinfo {author} {\bibfnamefont
			{P.}~\bibnamefont {Wu}}, \bibinfo {author} {\bibfnamefont {F.}~\bibnamefont
			{Yu}}, \bibinfo {author} {\bibfnamefont {J.}~\bibnamefont {Li}}, \bibinfo
		{author} {\bibfnamefont {M.}~\bibnamefont {Shan}}, \bibinfo {author}
		{\bibfnamefont {D.}~\bibnamefont {Zhao}}, \bibinfo {author} {\bibfnamefont
			{S.}~\bibnamefont {Li}}, \bibinfo {author} {\bibfnamefont {B.}~\bibnamefont
			{Kang}}, \bibinfo {author} {\bibfnamefont {Z.}~\bibnamefont {Wu}}, \bibinfo
		{author} {\bibfnamefont {Y.}~\bibnamefont {Zhou}}, \bibinfo {author}
		{\bibfnamefont {K.}~\bibnamefont {Liu}}, \bibinfo {author} {\bibfnamefont
			{Z.}~\bibnamefont {Xiang}}, \bibinfo {author} {\bibfnamefont
			{J.}~\bibnamefont {Ying}}, \bibinfo {author} {\bibfnamefont {Z.}~\bibnamefont
			{Wang}}, \bibinfo {author} {\bibfnamefont {T.}~\bibnamefont {Wu}}, \ and\
		\bibinfo {author} {\bibfnamefont {X.}~\bibnamefont {Chen}},\ }\href@noop {}
	{\bibfield  {journal} {\bibinfo  {journal} {Nature}\ }\textbf {\bibinfo
			{volume} {604}},\ \bibinfo {pages} {59} (\bibinfo {year} {2022})}\BibitemShut
	{NoStop}%
	\bibitem [{\citenamefont {Wei}\ \emph {et~al.}(2024)\citenamefont {Wei},
		\citenamefont {Tian}, \citenamefont {Cui}, \citenamefont {Zhai},
		\citenamefont {Li}, \citenamefont {Liu}, \citenamefont {Song}, \citenamefont
		{Feng}, \citenamefont {Huang}, \citenamefont {Wang} \emph
		{et~al.}}]{wei2024three}%
	\BibitemOpen
	\bibfield  {author} {\bibinfo {author} {\bibfnamefont {X.}~\bibnamefont
			{Wei}}, \bibinfo {author} {\bibfnamefont {C.}~\bibnamefont {Tian}}, \bibinfo
		{author} {\bibfnamefont {H.}~\bibnamefont {Cui}}, \bibinfo {author}
		{\bibfnamefont {Y.}~\bibnamefont {Zhai}}, \bibinfo {author} {\bibfnamefont
			{Y.}~\bibnamefont {Li}}, \bibinfo {author} {\bibfnamefont {S.}~\bibnamefont
			{Liu}}, \bibinfo {author} {\bibfnamefont {Y.}~\bibnamefont {Song}}, \bibinfo
		{author} {\bibfnamefont {Y.}~\bibnamefont {Feng}}, \bibinfo {author}
		{\bibfnamefont {M.}~\bibnamefont {Huang}}, \bibinfo {author} {\bibfnamefont
			{Z.}~\bibnamefont {Wang}},  \emph {et~al.},\ }\href@noop {} {\bibfield
		{journal} {\bibinfo  {journal} {Nature Communications}\ }\textbf {\bibinfo
			{volume} {15}},\ \bibinfo {pages} {5038} (\bibinfo {year}
		{2024})}\BibitemShut {NoStop}%
	\bibitem [{\citenamefont {Yin}\ \emph {et~al.}(2022)\citenamefont {Yin},
		\citenamefont {Lian},\ and\ \citenamefont {Hasan}}]{KagomeReview}%
	\BibitemOpen
	\bibfield  {author} {\bibinfo {author} {\bibfnamefont {J.-X.}\ \bibnamefont
			{Yin}}, \bibinfo {author} {\bibfnamefont {B.}~\bibnamefont {Lian}}, \ and\
		\bibinfo {author} {\bibfnamefont {M.~Z.}\ \bibnamefont {Hasan}},\ }\href@noop
	{} {\bibfield  {journal} {\bibinfo  {journal} {Nature}\ }\textbf {\bibinfo
			{volume} {612}},\ \bibinfo {pages} {647} (\bibinfo {year}
		{2022})}\BibitemShut {NoStop}%
	\bibitem [{\citenamefont {Neupert}\ \emph {et~al.}(2022)\citenamefont
		{Neupert}, \citenamefont {Denner}, \citenamefont {Yin}, \citenamefont
		{Thomale},\ and\ \citenamefont {Hasan}}]{Neupert2022}%
	\BibitemOpen
	\bibfield  {author} {\bibinfo {author} {\bibfnamefont {T.}~\bibnamefont
			{Neupert}}, \bibinfo {author} {\bibfnamefont {M.~M.}\ \bibnamefont {Denner}},
		\bibinfo {author} {\bibfnamefont {J.-X.}\ \bibnamefont {Yin}}, \bibinfo
		{author} {\bibfnamefont {R.}~\bibnamefont {Thomale}}, \ and\ \bibinfo
		{author} {\bibfnamefont {M.~Z.}\ \bibnamefont {Hasan}},\ }\href@noop {}
	{\bibfield  {journal} {\bibinfo  {journal} {Nature Physics}\ }\textbf
		{\bibinfo {volume} {18}},\ \bibinfo {pages} {137} (\bibinfo {year}
		{2022})}\BibitemShut {NoStop}%
	\bibitem [{\citenamefont {Christensen}\ \emph {et~al.}(2022)\citenamefont
		{Christensen}, \citenamefont {Birol}, \citenamefont {Andersen},\ and\
		\citenamefont {Fernandes}}]{Christensen2}%
	\BibitemOpen
	\bibfield  {author} {\bibinfo {author} {\bibfnamefont {M.~H.}\ \bibnamefont
			{Christensen}}, \bibinfo {author} {\bibfnamefont {T.}~\bibnamefont {Birol}},
		\bibinfo {author} {\bibfnamefont {B.~M.}\ \bibnamefont {Andersen}}, \ and\
		\bibinfo {author} {\bibfnamefont {R.~M.}\ \bibnamefont {Fernandes}},\
	}\href@noop {} {\bibfield  {journal} {\bibinfo  {journal} {Phys. Rev. B}\
		}\textbf {\bibinfo {volume} {106}},\ \bibinfo {pages} {144504} (\bibinfo
		{year} {2022})}\BibitemShut {NoStop}%
	\bibitem [{\citenamefont {Denner}\ \emph
		{et~al.}(2021{\natexlab{a}})\citenamefont {Denner}, \citenamefont {Thomale},\
		and\ \citenamefont {Neupert}}]{TitusAdd}%
	\BibitemOpen
	\bibfield  {author} {\bibinfo {author} {\bibfnamefont {M.~M.}\ \bibnamefont
			{Denner}}, \bibinfo {author} {\bibfnamefont {R.}~\bibnamefont {Thomale}}, \
		and\ \bibinfo {author} {\bibfnamefont {T.}~\bibnamefont {Neupert}},\
	}\href@noop {} {\bibfield  {journal} {\bibinfo  {journal} {Phys. Rev. Lett.}\
		}\textbf {\bibinfo {volume} {127}},\ \bibinfo {pages} {217601} (\bibinfo
		{year} {2021}{\natexlab{a}})}\BibitemShut {NoStop}%
	\bibitem [{\citenamefont {Tazai}\ \emph {et~al.}(2023)\citenamefont {Tazai},
		\citenamefont {Yamakawa},\ and\ \citenamefont {Kontani}}]{Tazai}%
	\BibitemOpen
	\bibfield  {author} {\bibinfo {author} {\bibfnamefont {R.}~\bibnamefont
			{Tazai}}, \bibinfo {author} {\bibfnamefont {Y.}~\bibnamefont {Yamakawa}}, \
		and\ \bibinfo {author} {\bibfnamefont {H.}~\bibnamefont {Kontani}},\
	}\href@noop {} {\bibfield  {journal} {\bibinfo  {journal} {arXiv:2303.00623}\
		} (\bibinfo {year} {2023})}\BibitemShut {NoStop}%
	\bibitem [{\citenamefont {Denner}\ \emph
		{et~al.}(2021{\natexlab{b}})\citenamefont {Denner}, \citenamefont {Thomale},\
		and\ \citenamefont {Neupert}}]{Denner}%
	\BibitemOpen
	\bibfield  {author} {\bibinfo {author} {\bibfnamefont {M.~M.}\ \bibnamefont
			{Denner}}, \bibinfo {author} {\bibfnamefont {R.}~\bibnamefont {Thomale}}, \
		and\ \bibinfo {author} {\bibfnamefont {T.}~\bibnamefont {Neupert}},\
	}\href@noop {} {\bibfield  {journal} {\bibinfo  {journal} {Phys. Rev. Lett.}\
		}\textbf {\bibinfo {volume} {127}},\ \bibinfo {pages} {217601} (\bibinfo
		{year} {2021}{\natexlab{b}})}\BibitemShut {NoStop}%
	\bibitem [{\citenamefont {Deng}\ \emph {et~al.}(2024)\citenamefont {Deng},
		\citenamefont {Qin}, \citenamefont {Liu}, \citenamefont {Yang}, \citenamefont
		{Fu}, \citenamefont {Zhang}, \citenamefont {Wu}, \citenamefont {Wang},
		\citenamefont {Shi}, \citenamefont {Liu}, \citenamefont {Liu}, \citenamefont
		{Yan}, \citenamefont {Song}, \citenamefont {Xu}, \citenamefont {Zhao},
		\citenamefont {Yi}, \citenamefont {Xu}, \citenamefont {Hohmann},
		\citenamefont {Holb{\ae}k}, \citenamefont {D{\"u}rrnagel}, \citenamefont
		{Zhou}, \citenamefont {Chang}, \citenamefont {Yao}, \citenamefont {Wang},
		\citenamefont {Guguchia}, \citenamefont {Neupert}, \citenamefont {Thomale},
		\citenamefont {Fischer},\ and\ \citenamefont {Yin}}]{deng2024chiral}%
	\BibitemOpen
	\bibfield  {author} {\bibinfo {author} {\bibfnamefont {H.}~\bibnamefont
			{Deng}}, \bibinfo {author} {\bibfnamefont {H.}~\bibnamefont {Qin}}, \bibinfo
		{author} {\bibfnamefont {G.}~\bibnamefont {Liu}}, \bibinfo {author}
		{\bibfnamefont {T.}~\bibnamefont {Yang}}, \bibinfo {author} {\bibfnamefont
			{R.}~\bibnamefont {Fu}}, \bibinfo {author} {\bibfnamefont {Z.}~\bibnamefont
			{Zhang}}, \bibinfo {author} {\bibfnamefont {X.}~\bibnamefont {Wu}}, \bibinfo
		{author} {\bibfnamefont {Z.}~\bibnamefont {Wang}}, \bibinfo {author}
		{\bibfnamefont {Y.}~\bibnamefont {Shi}}, \bibinfo {author} {\bibfnamefont
			{J.}~\bibnamefont {Liu}}, \bibinfo {author} {\bibfnamefont {H.}~\bibnamefont
			{Liu}}, \bibinfo {author} {\bibfnamefont {X.-Y.}\ \bibnamefont {Yan}},
		\bibinfo {author} {\bibfnamefont {W.}~\bibnamefont {Song}}, \bibinfo {author}
		{\bibfnamefont {X.}~\bibnamefont {Xu}}, \bibinfo {author} {\bibfnamefont
			{Y.}~\bibnamefont {Zhao}}, \bibinfo {author} {\bibfnamefont {M.}~\bibnamefont
			{Yi}}, \bibinfo {author} {\bibfnamefont {G.}~\bibnamefont {Xu}}, \bibinfo
		{author} {\bibfnamefont {H.}~\bibnamefont {Hohmann}}, \bibinfo {author}
		{\bibfnamefont {S.~C.}\ \bibnamefont {Holb{\ae}k}}, \bibinfo {author}
		{\bibfnamefont {M.}~\bibnamefont {D{\"u}rrnagel}}, \bibinfo {author}
		{\bibfnamefont {S.}~\bibnamefont {Zhou}}, \bibinfo {author} {\bibfnamefont
			{G.}~\bibnamefont {Chang}}, \bibinfo {author} {\bibfnamefont
			{Y.}~\bibnamefont {Yao}}, \bibinfo {author} {\bibfnamefont {Q.}~\bibnamefont
			{Wang}}, \bibinfo {author} {\bibfnamefont {Z.}~\bibnamefont {Guguchia}},
		\bibinfo {author} {\bibfnamefont {T.}~\bibnamefont {Neupert}}, \bibinfo
		{author} {\bibfnamefont {R.}~\bibnamefont {Thomale}}, \bibinfo {author}
		{\bibfnamefont {M.~H.}\ \bibnamefont {Fischer}}, \ and\ \bibinfo {author}
		{\bibfnamefont {J.-X.}\ \bibnamefont {Yin}},\ }\href@noop {} {\bibfield
		{journal} {\bibinfo  {journal} {Nature}\ }\textbf {\bibinfo {volume} {632}},\
		\bibinfo {pages} {775} (\bibinfo {year} {2024})}\BibitemShut {NoStop}%
	\bibitem [{\citenamefont {Scammell}\ \emph {et~al.}(2023)\citenamefont
		{Scammell}, \citenamefont {Ingham}, \citenamefont {Li},\ and\ \citenamefont
		{Sushkov}}]{scammell2023chiral}%
	\BibitemOpen
	\bibfield  {author} {\bibinfo {author} {\bibfnamefont {H.~D.}\ \bibnamefont
			{Scammell}}, \bibinfo {author} {\bibfnamefont {J.}~\bibnamefont {Ingham}},
		\bibinfo {author} {\bibfnamefont {T.}~\bibnamefont {Li}}, \ and\ \bibinfo
		{author} {\bibfnamefont {O.~P.}\ \bibnamefont {Sushkov}},\ }\href@noop {}
	{\bibfield  {journal} {\bibinfo  {journal} {Nature Communications}\ }\textbf
		{\bibinfo {volume} {14}},\ \bibinfo {pages} {605} (\bibinfo {year}
		{2023})}\BibitemShut {NoStop}%
	\bibitem [{\citenamefont {Ingham}\ \emph {et~al.}(2025)\citenamefont {Ingham},
		\citenamefont {Thomale},\ and\ \citenamefont
		{Scammell}}]{ingham2025vestigial}%
	\BibitemOpen
	\bibfield  {author} {\bibinfo {author} {\bibfnamefont {J.}~\bibnamefont
			{Ingham}}, \bibinfo {author} {\bibfnamefont {R.}~\bibnamefont {Thomale}}, \
		and\ \bibinfo {author} {\bibfnamefont {H.~D.}\ \bibnamefont {Scammell}},\
	}\href@noop {} {\bibfield  {journal} {\bibinfo  {journal} {arXiv preprint
				arXiv:2503.02929}\ } (\bibinfo {year} {2025})}\BibitemShut {NoStop}%
	\bibitem [{\citenamefont {Guo}\ \emph {et~al.}(2024{\natexlab{b}})\citenamefont
		{Guo}, \citenamefont {van Delft}, \citenamefont {Gutierrez-Amigo},
		\citenamefont {Chen}, \citenamefont {Putzke}, \citenamefont {Wagner},
		\citenamefont {Fischer}, \citenamefont {Neupert}, \citenamefont {Errea},
		\citenamefont {Vergniory} \emph {et~al.}}]{guo2024distinct}%
	\BibitemOpen
	\bibfield  {author} {\bibinfo {author} {\bibfnamefont {C.}~\bibnamefont
			{Guo}}, \bibinfo {author} {\bibfnamefont {M.~R.}\ \bibnamefont {van Delft}},
		\bibinfo {author} {\bibfnamefont {M.}~\bibnamefont {Gutierrez-Amigo}},
		\bibinfo {author} {\bibfnamefont {D.}~\bibnamefont {Chen}}, \bibinfo {author}
		{\bibfnamefont {C.}~\bibnamefont {Putzke}}, \bibinfo {author} {\bibfnamefont
			{G.}~\bibnamefont {Wagner}}, \bibinfo {author} {\bibfnamefont {M.~H.}\
			\bibnamefont {Fischer}}, \bibinfo {author} {\bibfnamefont {T.}~\bibnamefont
			{Neupert}}, \bibinfo {author} {\bibfnamefont {I.}~\bibnamefont {Errea}},
		\bibinfo {author} {\bibfnamefont {M.~G.}\ \bibnamefont {Vergniory}},  \emph
		{et~al.},\ }\href@noop {} {\bibfield  {journal} {\bibinfo  {journal} {npj
				Quantum Materials}\ }\textbf {\bibinfo {volume} {9}},\ \bibinfo {pages} {20}
		(\bibinfo {year} {2024}{\natexlab{b}})}\BibitemShut {NoStop}%
	\bibitem [{\citenamefont {Qian}\ \emph {et~al.}(2021)\citenamefont {Qian},
		\citenamefont {Christensen}, \citenamefont {Hu}, \citenamefont {Saha},
		\citenamefont {Andersen}, \citenamefont {Fernandes}, \citenamefont {Birol},\
		and\ \citenamefont {Ni}}]{Strain}%
	\BibitemOpen
	\bibfield  {author} {\bibinfo {author} {\bibfnamefont {T.}~\bibnamefont
			{Qian}}, \bibinfo {author} {\bibfnamefont {M.~H.}\ \bibnamefont
			{Christensen}}, \bibinfo {author} {\bibfnamefont {C.}~\bibnamefont {Hu}},
		\bibinfo {author} {\bibfnamefont {A.}~\bibnamefont {Saha}}, \bibinfo {author}
		{\bibfnamefont {B.~M.}\ \bibnamefont {Andersen}}, \bibinfo {author}
		{\bibfnamefont {R.~M.}\ \bibnamefont {Fernandes}}, \bibinfo {author}
		{\bibfnamefont {T.}~\bibnamefont {Birol}}, \ and\ \bibinfo {author}
		{\bibfnamefont {N.}~\bibnamefont {Ni}},\ }\href@noop {} {\bibfield  {journal}
		{\bibinfo  {journal} {Phys. Rev. B}\ }\textbf {\bibinfo {volume} {104}},\
		\bibinfo {pages} {144506} (\bibinfo {year} {2021})}\BibitemShut {NoStop}%
	\bibitem [{\citenamefont {Bachmann}\ \emph
		{et~al.}(2019{\natexlab{b}})\citenamefont {Bachmann}, \citenamefont
		{Ferguson}, \citenamefont {Theuss}, \citenamefont {Meng}, \citenamefont
		{Putzke}, \citenamefont {Helm}, \citenamefont {Shirer}, \citenamefont {Li},
		\citenamefont {Modic}, \citenamefont {Nicklas}, \citenamefont {König},
		\citenamefont {Low}, \citenamefont {Ghosh}, \citenamefont {Mackenzie},
		\citenamefont {Arnold}, \citenamefont {Hassinger}, \citenamefont {McDonald},
		\citenamefont {Winter}, \citenamefont {Bauer}, \citenamefont {Ronning},
		\citenamefont {Ramshaw}, \citenamefont {Nowack},\ and\ \citenamefont
		{Moll}}]{Maja}%
	\BibitemOpen
	\bibfield  {author} {\bibinfo {author} {\bibfnamefont {M.~D.}\ \bibnamefont
			{Bachmann}}, \bibinfo {author} {\bibfnamefont {G.~M.}\ \bibnamefont
			{Ferguson}}, \bibinfo {author} {\bibfnamefont {F.}~\bibnamefont {Theuss}},
		\bibinfo {author} {\bibfnamefont {T.}~\bibnamefont {Meng}}, \bibinfo {author}
		{\bibfnamefont {C.}~\bibnamefont {Putzke}}, \bibinfo {author} {\bibfnamefont
			{T.}~\bibnamefont {Helm}}, \bibinfo {author} {\bibfnamefont {K.~R.}\
			\bibnamefont {Shirer}}, \bibinfo {author} {\bibfnamefont {Y.-S.}\
			\bibnamefont {Li}}, \bibinfo {author} {\bibfnamefont {K.~A.}\ \bibnamefont
			{Modic}}, \bibinfo {author} {\bibfnamefont {M.}~\bibnamefont {Nicklas}},
		\bibinfo {author} {\bibfnamefont {M.}~\bibnamefont {König}}, \bibinfo
		{author} {\bibfnamefont {D.}~\bibnamefont {Low}}, \bibinfo {author}
		{\bibfnamefont {S.}~\bibnamefont {Ghosh}}, \bibinfo {author} {\bibfnamefont
			{A.~P.}\ \bibnamefont {Mackenzie}}, \bibinfo {author} {\bibfnamefont
			{F.}~\bibnamefont {Arnold}}, \bibinfo {author} {\bibfnamefont
			{E.}~\bibnamefont {Hassinger}}, \bibinfo {author} {\bibfnamefont {R.~D.}\
			\bibnamefont {McDonald}}, \bibinfo {author} {\bibfnamefont {L.~E.}\
			\bibnamefont {Winter}}, \bibinfo {author} {\bibfnamefont {E.~D.}\
			\bibnamefont {Bauer}}, \bibinfo {author} {\bibfnamefont {F.}~\bibnamefont
			{Ronning}}, \bibinfo {author} {\bibfnamefont {B.~J.}\ \bibnamefont
			{Ramshaw}}, \bibinfo {author} {\bibfnamefont {K.~C.}\ \bibnamefont {Nowack}},
		\ and\ \bibinfo {author} {\bibfnamefont {P.~J.~W.}\ \bibnamefont {Moll}},\
	}\href@noop {} {\bibfield  {journal} {\bibinfo  {journal} {Science}\ }\textbf
		{\bibinfo {volume} {366}},\ \bibinfo {pages} {221} (\bibinfo {year}
		{2019}{\natexlab{b}})}\BibitemShut {NoStop}%
	\bibitem [{\citenamefont {van Delft}\ \emph {et~al.}(2022)\citenamefont {van
			Delft}, \citenamefont {Bachmann}, \citenamefont {Putzke}, \citenamefont
		{Guo}, \citenamefont {Straquadine}, \citenamefont {Bauer}, \citenamefont
		{Ronning},\ and\ \citenamefont {Moll}}]{Maarten}%
	\BibitemOpen
	\bibfield  {author} {\bibinfo {author} {\bibfnamefont {M.~R.}\ \bibnamefont
			{van Delft}}, \bibinfo {author} {\bibfnamefont {M.~D.}\ \bibnamefont
			{Bachmann}}, \bibinfo {author} {\bibfnamefont {C.}~\bibnamefont {Putzke}},
		\bibinfo {author} {\bibfnamefont {C.}~\bibnamefont {Guo}}, \bibinfo {author}
		{\bibfnamefont {J.~A.~W.}\ \bibnamefont {Straquadine}}, \bibinfo {author}
		{\bibfnamefont {E.~D.}\ \bibnamefont {Bauer}}, \bibinfo {author}
		{\bibfnamefont {F.}~\bibnamefont {Ronning}}, \ and\ \bibinfo {author}
		{\bibfnamefont {P.~J.~W.}\ \bibnamefont {Moll}},\ }\href@noop {} {\bibfield
		{journal} {\bibinfo  {journal} {Applied Physics Letters}\ }\textbf {\bibinfo
			{volume} {120}},\ \bibinfo {pages} {092601} (\bibinfo {year}
		{2022})}\BibitemShut {NoStop}%
	\bibitem [{\citenamefont {Zhao}\ \emph {et~al.}(2021)\citenamefont {Zhao},
		\citenamefont {Li}, \citenamefont {Ortiz}, \citenamefont {Teicher},
		\citenamefont {Park}, \citenamefont {Ye}, \citenamefont {Wang}, \citenamefont
		{Balents}, \citenamefont {Wilson},\ and\ \citenamefont
		{Zeljkovic}}]{zhao2021cascade}%
	\BibitemOpen
	\bibfield  {author} {\bibinfo {author} {\bibfnamefont {H.}~\bibnamefont
			{Zhao}}, \bibinfo {author} {\bibfnamefont {H.}~\bibnamefont {Li}}, \bibinfo
		{author} {\bibfnamefont {B.~R.}\ \bibnamefont {Ortiz}}, \bibinfo {author}
		{\bibfnamefont {S.~M.}\ \bibnamefont {Teicher}}, \bibinfo {author}
		{\bibfnamefont {T.}~\bibnamefont {Park}}, \bibinfo {author} {\bibfnamefont
			{M.}~\bibnamefont {Ye}}, \bibinfo {author} {\bibfnamefont {Z.}~\bibnamefont
			{Wang}}, \bibinfo {author} {\bibfnamefont {L.}~\bibnamefont {Balents}},
		\bibinfo {author} {\bibfnamefont {S.~D.}\ \bibnamefont {Wilson}}, \ and\
		\bibinfo {author} {\bibfnamefont {I.}~\bibnamefont {Zeljkovic}},\ }\href@noop
	{} {\bibfield  {journal} {\bibinfo  {journal} {Nature}\ }\textbf {\bibinfo
			{volume} {599}},\ \bibinfo {pages} {216} (\bibinfo {year}
		{2021})}\BibitemShut {NoStop}%
	\bibitem [{\citenamefont {Huang}\ \emph {et~al.}(2022)\citenamefont {Huang},
		\citenamefont {Guo}, \citenamefont {Putzke}, \citenamefont {Gutierrez-Amigo},
		\citenamefont {Sun}, \citenamefont {Vergniory}, \citenamefont {Errea},
		\citenamefont {Chen}, \citenamefont {Felser},\ and\ \citenamefont
		{Moll}}]{huang2022mixed}%
	\BibitemOpen
	\bibfield  {author} {\bibinfo {author} {\bibfnamefont {X.}~\bibnamefont
			{Huang}}, \bibinfo {author} {\bibfnamefont {C.}~\bibnamefont {Guo}}, \bibinfo
		{author} {\bibfnamefont {C.}~\bibnamefont {Putzke}}, \bibinfo {author}
		{\bibfnamefont {M.}~\bibnamefont {Gutierrez-Amigo}}, \bibinfo {author}
		{\bibfnamefont {Y.}~\bibnamefont {Sun}}, \bibinfo {author} {\bibfnamefont
			{M.~G.}\ \bibnamefont {Vergniory}}, \bibinfo {author} {\bibfnamefont
			{I.}~\bibnamefont {Errea}}, \bibinfo {author} {\bibfnamefont
			{D.}~\bibnamefont {Chen}}, \bibinfo {author} {\bibfnamefont {C.}~\bibnamefont
			{Felser}}, \ and\ \bibinfo {author} {\bibfnamefont {P.~J.~W.}\ \bibnamefont
			{Moll}},\ }\href@noop {} {\bibfield  {journal} {\bibinfo  {journal} {Phys.
				Rev. B}\ }\textbf {\bibinfo {volume} {106}},\ \bibinfo {pages} {064510}
		(\bibinfo {year} {2022})}\BibitemShut {NoStop}%
	\bibitem [{\citenamefont {Ortiz}\ \emph {et~al.}(2021)\citenamefont {Ortiz},
		\citenamefont {Teicher}, \citenamefont {Kautzsch}, \citenamefont {Sarte},
		\citenamefont {Ruff}, \citenamefont {Seshadri},\ and\ \citenamefont
		{Wilson}}]{ortiz2021fermi}%
	\BibitemOpen
	\bibfield  {author} {\bibinfo {author} {\bibfnamefont {B.~R.}\ \bibnamefont
			{Ortiz}}, \bibinfo {author} {\bibfnamefont {S.~M.}\ \bibnamefont {Teicher}},
		\bibinfo {author} {\bibfnamefont {L.}~\bibnamefont {Kautzsch}}, \bibinfo
		{author} {\bibfnamefont {P.~M.}\ \bibnamefont {Sarte}}, \bibinfo {author}
		{\bibfnamefont {J.~P.}\ \bibnamefont {Ruff}}, \bibinfo {author}
		{\bibfnamefont {R.}~\bibnamefont {Seshadri}}, \ and\ \bibinfo {author}
		{\bibfnamefont {S.~D.}\ \bibnamefont {Wilson}},\ }\href@noop {} {\bibfield
		{journal} {\bibinfo  {journal} {Phys. Rev. X}\ }\textbf {\bibinfo {volume}
			{11}},\ \bibinfo {pages} {041030} (\bibinfo {year} {2021})}\BibitemShut
	{NoStop}%
	\bibitem [{\citenamefont {Chen}\ \emph {et~al.}(2023)\citenamefont {Chen},
		\citenamefont {Zheng}, \citenamefont {Zhang}, \citenamefont {Chan},
		\citenamefont {Zhu}, \citenamefont {Jenkins}, \citenamefont {Yu},
		\citenamefont {Shi}, \citenamefont {Ying}, \citenamefont {Xiang} \emph
		{et~al.}}]{chen2023magnetic}%
	\BibitemOpen
	\bibfield  {author} {\bibinfo {author} {\bibfnamefont {K.-W.}\ \bibnamefont
			{Chen}}, \bibinfo {author} {\bibfnamefont {G.}~\bibnamefont {Zheng}},
		\bibinfo {author} {\bibfnamefont {D.}~\bibnamefont {Zhang}}, \bibinfo
		{author} {\bibfnamefont {A.}~\bibnamefont {Chan}}, \bibinfo {author}
		{\bibfnamefont {Y.}~\bibnamefont {Zhu}}, \bibinfo {author} {\bibfnamefont
			{K.}~\bibnamefont {Jenkins}}, \bibinfo {author} {\bibfnamefont
			{F.}~\bibnamefont {Yu}}, \bibinfo {author} {\bibfnamefont {M.}~\bibnamefont
			{Shi}}, \bibinfo {author} {\bibfnamefont {J.}~\bibnamefont {Ying}}, \bibinfo
		{author} {\bibfnamefont {Z.}~\bibnamefont {Xiang}},  \emph {et~al.},\
	}\href@noop {} {\bibfield  {journal} {\bibinfo  {journal} {Communications
				Materials}\ }\textbf {\bibinfo {volume} {4}},\ \bibinfo {pages} {96}
		(\bibinfo {year} {2023})}\BibitemShut {NoStop}%
	\bibitem [{\citenamefont {Chapai}\ \emph {et~al.}(2023)\citenamefont {Chapai},
		\citenamefont {Leroux}, \citenamefont {Oliviero}, \citenamefont {Vignolles},
		\citenamefont {Bruyant}, \citenamefont {Smylie}, \citenamefont {Chung},
		\citenamefont {Kanatzidis}, \citenamefont {Kwok}, \citenamefont {Mitchell}
		\emph {et~al.}}]{chapai2023magnetic}%
	\BibitemOpen
	\bibfield  {author} {\bibinfo {author} {\bibfnamefont {R.}~\bibnamefont
			{Chapai}}, \bibinfo {author} {\bibfnamefont {M.}~\bibnamefont {Leroux}},
		\bibinfo {author} {\bibfnamefont {V.}~\bibnamefont {Oliviero}}, \bibinfo
		{author} {\bibfnamefont {D.}~\bibnamefont {Vignolles}}, \bibinfo {author}
		{\bibfnamefont {N.}~\bibnamefont {Bruyant}}, \bibinfo {author} {\bibfnamefont
			{M.}~\bibnamefont {Smylie}}, \bibinfo {author} {\bibfnamefont
			{D.}~\bibnamefont {Chung}}, \bibinfo {author} {\bibfnamefont
			{M.}~\bibnamefont {Kanatzidis}}, \bibinfo {author} {\bibfnamefont {W.-K.}\
			\bibnamefont {Kwok}}, \bibinfo {author} {\bibfnamefont {J.}~\bibnamefont
			{Mitchell}},  \emph {et~al.},\ }\href@noop {} {\bibfield  {journal} {\bibinfo
			{journal} {Physical review letters}\ }\textbf {\bibinfo {volume} {130}},\
		\bibinfo {pages} {126401} (\bibinfo {year} {2023})}\BibitemShut {NoStop}%
	\bibitem [{\citenamefont {Xie}\ \emph {et~al.}(2022{\natexlab{b}})\citenamefont
		{Xie}, \citenamefont {Li}, \citenamefont {Bourges}, \citenamefont {Ivanov},
		\citenamefont {Ye}, \citenamefont {Yin}, \citenamefont {Hasan}, \citenamefont
		{Luo}, \citenamefont {Yao}, \citenamefont {Wang} \emph
		{et~al.}}]{xie2022electron}%
	\BibitemOpen
	\bibfield  {author} {\bibinfo {author} {\bibfnamefont {Y.}~\bibnamefont
			{Xie}}, \bibinfo {author} {\bibfnamefont {Y.}~\bibnamefont {Li}}, \bibinfo
		{author} {\bibfnamefont {P.}~\bibnamefont {Bourges}}, \bibinfo {author}
		{\bibfnamefont {A.}~\bibnamefont {Ivanov}}, \bibinfo {author} {\bibfnamefont
			{Z.}~\bibnamefont {Ye}}, \bibinfo {author} {\bibfnamefont {J.-X.}\
			\bibnamefont {Yin}}, \bibinfo {author} {\bibfnamefont {M.~Z.}\ \bibnamefont
			{Hasan}}, \bibinfo {author} {\bibfnamefont {A.}~\bibnamefont {Luo}}, \bibinfo
		{author} {\bibfnamefont {Y.}~\bibnamefont {Yao}}, \bibinfo {author}
		{\bibfnamefont {Z.}~\bibnamefont {Wang}},  \emph {et~al.},\ }\href@noop {}
	{\bibfield  {journal} {\bibinfo  {journal} {Physical Review B}\ }\textbf
		{\bibinfo {volume} {105}},\ \bibinfo {pages} {L140501} (\bibinfo {year}
		{2022}{\natexlab{b}})}\BibitemShut {NoStop}%
	\bibitem [{\citenamefont {He}\ \emph {et~al.}(2024)\citenamefont {He},
		\citenamefont {Peis}, \citenamefont {Cuddy}, \citenamefont {Zhao},
		\citenamefont {Li}, \citenamefont {Zhang}, \citenamefont {Stumberger},
		\citenamefont {Moritz}, \citenamefont {Yang}, \citenamefont {Gao} \emph
		{et~al.}}]{he2024anharmonic}%
	\BibitemOpen
	\bibfield  {author} {\bibinfo {author} {\bibfnamefont {G.}~\bibnamefont
			{He}}, \bibinfo {author} {\bibfnamefont {L.}~\bibnamefont {Peis}}, \bibinfo
		{author} {\bibfnamefont {E.~F.}\ \bibnamefont {Cuddy}}, \bibinfo {author}
		{\bibfnamefont {Z.}~\bibnamefont {Zhao}}, \bibinfo {author} {\bibfnamefont
			{D.}~\bibnamefont {Li}}, \bibinfo {author} {\bibfnamefont {Y.}~\bibnamefont
			{Zhang}}, \bibinfo {author} {\bibfnamefont {R.}~\bibnamefont {Stumberger}},
		\bibinfo {author} {\bibfnamefont {B.}~\bibnamefont {Moritz}}, \bibinfo
		{author} {\bibfnamefont {H.}~\bibnamefont {Yang}}, \bibinfo {author}
		{\bibfnamefont {H.}~\bibnamefont {Gao}},  \emph {et~al.},\ }\href@noop {}
	{\bibfield  {journal} {\bibinfo  {journal} {Nature Communications}\ }\textbf
		{\bibinfo {volume} {15}},\ \bibinfo {pages} {1895} (\bibinfo {year}
		{2024})}\BibitemShut {NoStop}%
	\bibitem [{\citenamefont {Hu}\ \emph {et~al.}(2022)\citenamefont {Hu},
		\citenamefont {Wu}, \citenamefont {Ortiz}, \citenamefont {Ju}, \citenamefont
		{Han}, \citenamefont {Ma}, \citenamefont {Plumb}, \citenamefont {Radovic},
		\citenamefont {Thomale}, \citenamefont {Wilson} \emph {et~al.}}]{hu2022rich}%
	\BibitemOpen
	\bibfield  {author} {\bibinfo {author} {\bibfnamefont {Y.}~\bibnamefont
			{Hu}}, \bibinfo {author} {\bibfnamefont {X.}~\bibnamefont {Wu}}, \bibinfo
		{author} {\bibfnamefont {B.~R.}\ \bibnamefont {Ortiz}}, \bibinfo {author}
		{\bibfnamefont {S.}~\bibnamefont {Ju}}, \bibinfo {author} {\bibfnamefont
			{X.}~\bibnamefont {Han}}, \bibinfo {author} {\bibfnamefont {J.}~\bibnamefont
			{Ma}}, \bibinfo {author} {\bibfnamefont {N.~C.}\ \bibnamefont {Plumb}},
		\bibinfo {author} {\bibfnamefont {M.}~\bibnamefont {Radovic}}, \bibinfo
		{author} {\bibfnamefont {R.}~\bibnamefont {Thomale}}, \bibinfo {author}
		{\bibfnamefont {S.~D.}\ \bibnamefont {Wilson}},  \emph {et~al.},\ }\href@noop
	{} {\bibfield  {journal} {\bibinfo  {journal} {Nature Communications}\
		}\textbf {\bibinfo {volume} {13}},\ \bibinfo {pages} {2220} (\bibinfo {year}
		{2022})}\BibitemShut {NoStop}%
	\bibitem [{\citenamefont {Kang}\ \emph {et~al.}(2022)\citenamefont {Kang},
		\citenamefont {Fang}, \citenamefont {Kim}, \citenamefont {Ortiz},
		\citenamefont {Ryu}, \citenamefont {Kim}, \citenamefont {Yoo}, \citenamefont
		{Sangiovanni}, \citenamefont {Di~Sante}, \citenamefont {Park}, \citenamefont
		{Jozwiak}, \citenamefont {Bostwick}, \citenamefont {Rotenberg}, \citenamefont
		{Kaxiras}, \citenamefont {Wilson}, \citenamefont {Park},\ and\ \citenamefont
		{Comin}}]{kang2022}%
	\BibitemOpen
	\bibfield  {author} {\bibinfo {author} {\bibfnamefont {M.}~\bibnamefont
			{Kang}}, \bibinfo {author} {\bibfnamefont {S.}~\bibnamefont {Fang}}, \bibinfo
		{author} {\bibfnamefont {J.-K.}\ \bibnamefont {Kim}}, \bibinfo {author}
		{\bibfnamefont {B.~R.}\ \bibnamefont {Ortiz}}, \bibinfo {author}
		{\bibfnamefont {S.~H.}\ \bibnamefont {Ryu}}, \bibinfo {author} {\bibfnamefont
			{J.}~\bibnamefont {Kim}}, \bibinfo {author} {\bibfnamefont {J.}~\bibnamefont
			{Yoo}}, \bibinfo {author} {\bibfnamefont {G.}~\bibnamefont {Sangiovanni}},
		\bibinfo {author} {\bibfnamefont {D.}~\bibnamefont {Di~Sante}}, \bibinfo
		{author} {\bibfnamefont {B.-G.}\ \bibnamefont {Park}}, \bibinfo {author}
		{\bibfnamefont {C.}~\bibnamefont {Jozwiak}}, \bibinfo {author} {\bibfnamefont
			{A.}~\bibnamefont {Bostwick}}, \bibinfo {author} {\bibfnamefont
			{E.}~\bibnamefont {Rotenberg}}, \bibinfo {author} {\bibfnamefont
			{E.}~\bibnamefont {Kaxiras}}, \bibinfo {author} {\bibfnamefont {S.~D.}\
			\bibnamefont {Wilson}}, \bibinfo {author} {\bibfnamefont {J.-H.}\
			\bibnamefont {Park}}, \ and\ \bibinfo {author} {\bibfnamefont
			{R.}~\bibnamefont {Comin}},\ }\href@noop {} {\bibfield  {journal} {\bibinfo
			{journal} {Nature Physics}\ }\textbf {\bibinfo {volume} {18}},\ \bibinfo
		{pages} {301} (\bibinfo {year} {2022})}\BibitemShut {NoStop}%
	\bibitem [{\citenamefont {Vilkelis}\ \emph {et~al.}(2023)\citenamefont
		{Vilkelis}, \citenamefont {Wang},\ and\ \citenamefont
		{Akhmerov}}]{vilkelis2023bloch}%
	\BibitemOpen
	\bibfield  {author} {\bibinfo {author} {\bibfnamefont {K.}~\bibnamefont
			{Vilkelis}}, \bibinfo {author} {\bibfnamefont {L.}~\bibnamefont {Wang}}, \
		and\ \bibinfo {author} {\bibfnamefont {A.~R.}\ \bibnamefont {Akhmerov}},\
	}\href@noop {} {\bibfield  {journal} {\bibinfo  {journal} {SciPost Physics}\
		}\textbf {\bibinfo {volume} {15}},\ \bibinfo {pages} {019} (\bibinfo {year}
		{2023})}\BibitemShut {NoStop}%
	\bibitem [{\citenamefont {Xing}\ \emph {et~al.}(2024)\citenamefont {Xing},
		\citenamefont {Bae}, \citenamefont {Ritz}, \citenamefont {Yang},
		\citenamefont {Birol}, \citenamefont {Capa~Salinas}, \citenamefont {Ortiz},
		\citenamefont {Wilson}, \citenamefont {Wang}, \citenamefont {Fernandes} \emph
		{et~al.}}]{xing2024optical}%
	\BibitemOpen
	\bibfield  {author} {\bibinfo {author} {\bibfnamefont {Y.}~\bibnamefont
			{Xing}}, \bibinfo {author} {\bibfnamefont {S.}~\bibnamefont {Bae}}, \bibinfo
		{author} {\bibfnamefont {E.}~\bibnamefont {Ritz}}, \bibinfo {author}
		{\bibfnamefont {F.}~\bibnamefont {Yang}}, \bibinfo {author} {\bibfnamefont
			{T.}~\bibnamefont {Birol}}, \bibinfo {author} {\bibfnamefont {A.~N.}\
			\bibnamefont {Capa~Salinas}}, \bibinfo {author} {\bibfnamefont {B.~R.}\
			\bibnamefont {Ortiz}}, \bibinfo {author} {\bibfnamefont {S.~D.}\ \bibnamefont
			{Wilson}}, \bibinfo {author} {\bibfnamefont {Z.}~\bibnamefont {Wang}},
		\bibinfo {author} {\bibfnamefont {R.~M.}\ \bibnamefont {Fernandes}},  \emph
		{et~al.},\ }\href@noop {} {\bibfield  {journal} {\bibinfo  {journal}
			{Nature}\ }\textbf {\bibinfo {volume} {631}},\ \bibinfo {pages} {60}
		(\bibinfo {year} {2024})}\BibitemShut {NoStop}%
	\bibitem [{\citenamefont {Jiang}\ \emph {et~al.}(2021)\citenamefont {Jiang},
		\citenamefont {Yin}, \citenamefont {Denner}, \citenamefont {Shumiya},
		\citenamefont {Ortiz}, \citenamefont {Xu}, \citenamefont {Guguchia},
		\citenamefont {He}, \citenamefont {Hossain}, \citenamefont {Liu},
		\citenamefont {Ruff}, \citenamefont {Kautzsch}, \citenamefont {Zhang},
		\citenamefont {Chang}, \citenamefont {Belopolski}, \citenamefont {Zhang},
		\citenamefont {Cochran}, \citenamefont {Multer}, \citenamefont {Litskevich},
		\citenamefont {Cheng}, \citenamefont {Yang}, \citenamefont {Wang},
		\citenamefont {Thomale}, \citenamefont {Neupert}, \citenamefont {Wilson},\
		and\ \citenamefont {Hasan}}]{Kchiral}%
	\BibitemOpen
	\bibfield  {author} {\bibinfo {author} {\bibfnamefont {Y.-X.}\ \bibnamefont
			{Jiang}}, \bibinfo {author} {\bibfnamefont {J.-X.}\ \bibnamefont {Yin}},
		\bibinfo {author} {\bibfnamefont {M.~M.}\ \bibnamefont {Denner}}, \bibinfo
		{author} {\bibfnamefont {N.}~\bibnamefont {Shumiya}}, \bibinfo {author}
		{\bibfnamefont {B.~R.}\ \bibnamefont {Ortiz}}, \bibinfo {author}
		{\bibfnamefont {G.}~\bibnamefont {Xu}}, \bibinfo {author} {\bibfnamefont
			{Z.}~\bibnamefont {Guguchia}}, \bibinfo {author} {\bibfnamefont
			{J.}~\bibnamefont {He}}, \bibinfo {author} {\bibfnamefont {M.~S.}\
			\bibnamefont {Hossain}}, \bibinfo {author} {\bibfnamefont {X.}~\bibnamefont
			{Liu}}, \bibinfo {author} {\bibfnamefont {J.}~\bibnamefont {Ruff}}, \bibinfo
		{author} {\bibfnamefont {L.}~\bibnamefont {Kautzsch}}, \bibinfo {author}
		{\bibfnamefont {S.~S.}\ \bibnamefont {Zhang}}, \bibinfo {author}
		{\bibfnamefont {G.}~\bibnamefont {Chang}}, \bibinfo {author} {\bibfnamefont
			{I.}~\bibnamefont {Belopolski}}, \bibinfo {author} {\bibfnamefont
			{Q.}~\bibnamefont {Zhang}}, \bibinfo {author} {\bibfnamefont {T.~A.}\
			\bibnamefont {Cochran}}, \bibinfo {author} {\bibfnamefont {D.}~\bibnamefont
			{Multer}}, \bibinfo {author} {\bibfnamefont {M.}~\bibnamefont {Litskevich}},
		\bibinfo {author} {\bibfnamefont {Z.-J.}\ \bibnamefont {Cheng}}, \bibinfo
		{author} {\bibfnamefont {X.~P.}\ \bibnamefont {Yang}}, \bibinfo {author}
		{\bibfnamefont {Z.}~\bibnamefont {Wang}}, \bibinfo {author} {\bibfnamefont
			{R.}~\bibnamefont {Thomale}}, \bibinfo {author} {\bibfnamefont
			{T.}~\bibnamefont {Neupert}}, \bibinfo {author} {\bibfnamefont {S.~D.}\
			\bibnamefont {Wilson}}, \ and\ \bibinfo {author} {\bibfnamefont {M.~Z.}\
			\bibnamefont {Hasan}},\ }\href@noop {} {\bibfield  {journal} {\bibinfo
			{journal} {Nature Materials}\ }\textbf {\bibinfo {volume} {20}},\ \bibinfo
		{pages} {1353} (\bibinfo {year} {2021})}\BibitemShut {NoStop}%
	\bibitem [{\citenamefont {Usui}\ \emph {et~al.}(2019)\citenamefont {Usui},
		\citenamefont {Ochi}, \citenamefont {Kitamura}, \citenamefont {Oka},
		\citenamefont {Ogura}, \citenamefont {Rosner}, \citenamefont {Haverkort},
		\citenamefont {Sunko}, \citenamefont {King}, \citenamefont {Mackenzie} \emph
		{et~al.}}]{usui2019hidden}%
	\BibitemOpen
	\bibfield  {author} {\bibinfo {author} {\bibfnamefont {H.}~\bibnamefont
			{Usui}}, \bibinfo {author} {\bibfnamefont {M.}~\bibnamefont {Ochi}}, \bibinfo
		{author} {\bibfnamefont {S.}~\bibnamefont {Kitamura}}, \bibinfo {author}
		{\bibfnamefont {T.}~\bibnamefont {Oka}}, \bibinfo {author} {\bibfnamefont
			{D.}~\bibnamefont {Ogura}}, \bibinfo {author} {\bibfnamefont
			{H.}~\bibnamefont {Rosner}}, \bibinfo {author} {\bibfnamefont {M.~W.}\
			\bibnamefont {Haverkort}}, \bibinfo {author} {\bibfnamefont {V.}~\bibnamefont
			{Sunko}}, \bibinfo {author} {\bibfnamefont {P.~D.}\ \bibnamefont {King}},
		\bibinfo {author} {\bibfnamefont {A.~P.}\ \bibnamefont {Mackenzie}},  \emph
		{et~al.},\ }\href@noop {} {\bibfield  {journal} {\bibinfo  {journal}
			{Physical Review Materials}\ }\textbf {\bibinfo {volume} {3}},\ \bibinfo
		{pages} {045002} (\bibinfo {year} {2019})}\BibitemShut {NoStop}%
\end{thebibliography}
\end{document}


\renewcommand{\theequation}{S\arabic{equation}}
	\newcommand{\beginMethods}{%
		\setcounter{table}{0}
		\renewcommand{\thetable}{S\arabic{table}}%
		\setcounter{figure}{0}
		\renewcommand{\thefigure}{S\arabic{figure}}%
	}

\title{Supplementary materials for "Long-range electron coherence in a Kagome metal"}

\author{Chunyu (Mark) Guo${}^{\dagger}$}\affiliation{Max Planck Institute for the Structure and Dynamics of Matter, Hamburg, Germany}
\author{Kaize Wang${}^{}$}
\affiliation{Max Planck Institute for the Structure and Dynamics of Matter, Hamburg, Germany}
\author{Ling Zhang${}^{}$}
\affiliation{Max Planck Institute for the Structure and Dynamics of Matter, Hamburg, Germany}
\author{Carsten Putzke${}^{}$}
\affiliation{Max Planck Institute for the Structure and Dynamics of Matter, Hamburg, Germany}
\author{Dong Chen}\affiliation{Max Planck Institute for Chemical Physics of Solids, Dresden, Germany}\affiliation{College of Physics, Qingdao University, Qingdao, China}
\author{Maarten R. van Delft}\affiliation{High Field Magnet Laboratory (HFML - EMFL), Radboud University, Toernooiveld 7, 6525 ED Nijmegen, The Netherlands}
\affiliation{Radboud University, Institute for Molecules and Materials, Nijmegen 6525 AJ, Netherlands}
\author{Steffen Wiedmann}\affiliation{High Field Magnet Laboratory (HFML - EMFL), Radboud University, Toernooiveld 7, 6525 ED Nijmegen, The Netherlands}
\affiliation{Radboud University, Institute for Molecules and Materials, Nijmegen 6525 AJ, Netherlands}
\author{Fedor F. Balakirev}\affiliation{National High Magnetic Field Laboratory, Los Alamos National Laboratory, Los Alamos, New Mexico 87545, USA}
\author{Ross McDonald}\affiliation{National High Magnetic Field Laboratory, Los Alamos National Laboratory, Los Alamos, New Mexico 87545, USA}
\author{Martin Gutierrez-Amigo}\affiliation{Department of Applied Physics, Aalto University School of Science, FI-00076 Aalto, Finland}
\author{Manex Alkorta}\affiliation{Centro de Física de Materiales (CSIC-UPV/EHU), Donostia-San Sebastian, Spain}
\affiliation{Department of Physics, University of the Basque Country (UPV/EHU), Bilbao, Spain}
\author{Ion Errea}\affiliation{Centro de Física de Materiales (CSIC-UPV/EHU), Donostia-San Sebastian, Spain}
\affiliation{Donostia International Physics Center, Donostia-San Sebastian, Spain}
\affiliation{Fisika Aplikatua Saila, Gipuzkoako Ingeniaritza Eskola, University of the Basque Country (UPV/EHU), Donostia-San Sebastian, Spain}
\author{Maia G. Vergniory}\affiliation{Donostia International Physics Center, Donostia-San Sebastian, Spain}
\affiliation{Max Planck Institute for Chemical Physics of Solids, Dresden, Germany}
\author{Takashi Oka}\affiliation{The Institute for Solid State Physics, The University of Tokyo, Kashiwa, Japan}
\author{Roderich Moessner}\affiliation{Max Planck Institute for the Physics of Complex Systems, Dresden, Germany}
\author{Mark H. Fischer}\affiliation{Department of Physics, University of Zürich, Zürich, Switzerland}
\author{Titus Neupert}\affiliation{Department of Physics, University of Zürich, Zürich, Switzerland}
\author{Claudia Felser}\affiliation{Max Planck Institute for Chemical Physics of Solids, Dresden, Germany}
\author{Philip J. W. Moll${}^{\dagger}$}\affiliation{Max Planck Institute for the Structure and Dynamics of Matter, Hamburg, Germany}

\date{\today}
\maketitle
\normalsize{$^\dagger$Corresponding authors: chunyu.guo@mpsd.mpg.de(C.G.); philip.moll@mpsd.mpg.de(P.J.W.M.).}

\subsection{Crystal synthesis and device fabrication}
Plate-like single crystals of CsV$_3$Sb$_5$ are obtained following a self-flux procedure as described in Ref. \cite{DresdenThermal}. They crystallize in the hexagonal structure (P6/mmm space group). The micro-devices (S1 to S8) are fabricated using the focused-ion-beam (FIB) technique (Fig. S \ref{SEM}). The microstructure fabrication procedure and the initial test on device quality are performed following the previously reported recipe\cite{Guo2022,guo2024correlated}. A slab of the bulk material (lamella) is dug out and transferred in situ by a micro-manipulator and welded to a gold-coated (Au:300~nm) SiN$_x$ membrane chip via Pt-deposition with ion beams. To clarify the influence of FIB-deposited contacting electrodes, device S3 is gold-coated after the lamella transfer process. The membrane window is later carved into meander-shaped springs, which become the only mechanical connections from the device to the supporting substrate. These soft springs with low spring constant $\approx$ 150~N/m significantly reduce the thermal contraction strain. The lamella is then fabricated into the desired shape following a low-voltage (2~kV) and low-current (100~pA) Xe ion beam cleaning as the final fabrication step to reduce the thickness of the FIB-damaged amorphous layer. To reduce the torque force at high magnetic fields, devices S3 and S5 are attached rigidly on one side of the silicon substrate with Pt-deposition, and the clear consistency among all devices demonstrates the irrelevance of the magnetic torque effect in our measurements. 

\subsection{Electrical measurements}
Resistance measurements were performed using a multichannel Synktek lockin in a Dynacool PPMS system with a maximal magnetic field of 14 T. A low AC current of ~30 $\mu$A is used to measure all devices to avoid Joule heating. The absence of notable self-heating has been confirmed by the absence of higher harmonic responses. The consistent observation of clear Shubnikov–de Haas oscillations and high electric conductivity at low temperatures demonstrates the intact physical properties of all FIB-fabricated devices and the irrelevance of the FIB-damaged amorphous layer, which typically encases the samples with a thickness of 10~nm corresponding to a negligible fraction of the sample cross-section. The field dependence of magnetoresistance at base temperature ($T$ = 2 K) shows a slight variation across different devices (Fig. S \ref{allSum}), mainly attributed to a small misalignment between the magnetic field direction and the Kagome plane. The subtracted second derivative demonstrates a consistent oscillation frequency change with varying sample width following the description of $h/e$ oscillations. For an estimation of uncertainty, we fit the width ($w$) dependence of the oscillation period ($\Delta B$) based on the relation: $\Phi_0=\Delta B w c'$. This yields $c' = 9.11 \pm 0.36~\textup{\AA}$, which is near identical to the $c$-axis lattice parameter in CsV$_3$Sb$_5$ ($c = 9.28 ~\textup{\AA}$). The errors of this analysis are mainly attributed to the uncertainty in determining the exact oscillation period $\Delta B$.

\subsection{Subtraction of oscillations: polynomial fitting versus second-order derivative}

Two independent methods have been used to remove the background of magnetoresistance to check the consistency of the $h/e$ oscillations subtracted (Fig. S \ref{Compare}). These methods follow the common practices of the analysis of SdH oscillations. First, we directly take the second-order derivative of the field dependence of the magnetoresistance, and the periodic-in-field oscillations are readily observable. When the magnetoresistance possesses a continuously varying second-order derivative, the resulting background may mask the oscillatory part of the signal. For the complimentary analysis, the MR is fitted with a 5th-order polynomial, and the subtracted part displays consistent oscillatory behavior. The polynomial was checked to ensure that it did not contain notable oscillatory behavior and did not introduce artificial oscillations into the data. This way, the background is subtracted further, while its direct comparison to the second-order derivative reveals a one-to-one correspondence with the pi-phase shift. It clearly demonstrates the consistency between these methods in analyzing $h/e$ oscillations. All raw data, as well as the background-subtracted data, are made available in the online repository.

\subsection{Influence of strain effect}

The electronic response in CsV$_3$Sb$_5$ is extremely sensitive to the influence of strain effect, as consistently demonstrated by previous observations\cite{Guo2022,guo2024correlated}. To explore the possible strain-sensitivity of $h/e$ oscillations, we mounted the device directly to a sapphire substrate without the membrane springs for mechanical buffering (Fig. S \ref{Strain}). The mismatch of the thermal expansion coefficient between the device and the sapphire substrate leads to a tensile strain on the device which reaches $\sim$ 0.3\% at $T$ = 2 K\cite{guo2024correlated}. Its impact is elaborated by the clear increase of the charge ordering temperature, while the broadening of the transition is due to the unavoidable strain inhomogeneity across the microstructure. Importantly, the $h/e$ oscillations are strongly suppressed in the strained device. Its comparison to a nearly strain-free device (S6) with a similar sample width demonstrates more than 90\% reduction of oscillation amplitude due to strain. Moreover, the oscillation disappears quickly with increasing field as only the first period can be clearly resolved. Given the actual change of lattice parameter due to tensile strain is less than 0.5\%, the deconstructive interference caused by the inhomogeneous lattice constant across the strained device cannot be the main source of the reduction observed. In contrast, it aligns with the observed suppression of strain in magneto-chiral transport signatures \cite{Guo2022}. This indicates a fundamental link between long-range coherence and correlated electronic order in CsV$_3$Sb$_5$.

\subsection{Transport mean free path analysis}

The transport mean free path $l_t$ denotes the distance an excited quasiparticle travels before its momentum has been randomized by scattering. The idea is based on the decay time of a DC transport current when the driving DC electric field is instantaneously removed at some point in time. Scattering processes with the lattice and its defects will transfer momentum from the electron system and eventually bring it to rest, leading to a characteristic current decay $I(t)=I_0 e^{-t/\tau_t}$. This transport lifetime $\tau_t$ can be translated to a mean distance between scattering events of a quasiparticle at the speed of the Fermi velocity $v_F$ simply via $l_t=v_F \tau_t$, which can be related to conductivities via Boltzmann transport. Such a picture of freely traveling quasiparticles between scattering events is critical in mesoscopic physics when device sizes are similar to this mean distance between scattering. However, this simple logic only rigorously works in simple cases, in particular in isotropic metals (spherical Fermi surface) with only one single band and carrier type. For complex multi-band metals hosting both electrons and holes on Fermi surface sheets with non-spherical shapes, this assumption is not generally valid. Scattering processes are momentum- and band-dependent, and a global decay time cannot be assumed to reflect directly onto each individual electron state. On top, the strong electron-phonon coupling results in non-trivial momentum dependences of the scattering matrix elements via the phonon-dispersion. As a result, the question of ballistic transport in such materials, in practice, is challenging. The goal is to apply the simplest analysis and show its mean free path to fall substantially below the device size, hence speaking against ballistic transport.

We start by estimating the transport mean free path based on the Drude model:
		\begin{equation}
\sigma_{xx} = n e^2 \tau / m^*
		\la{Drudeequ}
		\end{equation}
The conductivity is directly obtained from the measurements of in-plane resistivity of device S5. The carrier density is extracted from the previous report of Hall voltage measurements on CsV$_3$Sb$_5$ \cite{CVSHall}. At $T$ = 5 K, the hall resistivity $\rho_{xy}$ reaches 0.47 $\mu \Omega$ cm at $B$ = 14 T. Given the almost linear-in-field $\rho_{xy}$, the carrier density n is estimated to be $\approx 1.86$ x 10$^{22}$ cm$^{-3}$ under the assumption of single conduction band. For simplicity, we also assume that the carrier density is temperature-independent, as the temperature variation of Hall resistivity is mainly attributed to the evolution of carrier mobilities of different pockets while the electronic structure remains almost intact below 50 K. Since the Brillouin zone is majorly occupied by the hexagonal Fermi surface, the in-plane Fermi wavevector can be estimated by: $n = \frac{3 \sqrt{3}}{2} k_{F,x}^2 k_{F,z} / 4 \pi ^3$ with $k_{F,z}$ directly determined by the lattice constant $c$ $\approx$ 9$\textup{~\AA}$  ($k_{F,z} = 2 \pi /c $ ). Therefore, the transport mean free path can be calculated by:
		\begin{equation}
l_t = v_F \tau = \frac{\hbar k_{F,x} \sigma }{ n e^2} = \frac{\hbar \sigma}{e^2 \sqrt{0.1316 n /c}} 
		\la{lt}
		\end{equation}
which allows us to extract the transport mean free path directly from the temperature dependence of resistivity (Fig. S \ref{Drude}).  Right above the superconducting transition temperature ($T_{\text{c}}$ = 2.8 K), the transport mean free path reaches 500 nm and then quickly drops to 150 nm as the temperature increases to 20 K. These values are well below the characteristic sizes of all devices measured. 

As stated above, this estimate is based on crude approximations. Firstly, due to the slightly non-linear field dependence of Hall resistivity, the single-band assumption may lead to an underestimation of the total carrier count across multiple conduction bands. Secondly, the assumption of a singular, hexagonal 2D Fermi surface oversimplifies the complex Fermiology in CsV$_3$Sb$_5$, which affects the approximation of the Fermi wavelength based on carrier density. As the single-band assumption tends to overestimate the averaged Fermi velocity, these estimations are rather an upper bound for the transport mean free path $l_t$. Given the aforementioned difficulties, one may still argue for substantially suppressed scattering for some regions of one of the Fermi surfaces to reach a quasi-ballistic limit for these few select states only. However, this would imply an extreme variation of lifetime across the Fermi surface, a highly unusual feature of materials dominated by impurity scattering.

This further supports our conclusion that the sizes of the microstructures we studied are all substantially larger than $l_t$, especially at elevated temperatures.

\subsection{Dingle analysis for subtracting quantum mean free path}
The quantum mean free path is obtained by analyzing the field dependence of the SdH oscillation amplitude measured with the magnetic field applied along the $c$-direction (Fig. S \ref{mc}). According to the Lifshitz-Kosevich model, the oscillations can be described as\cite{guo2021temperature}:
		\begin{equation}
  \Delta\rho_{\parallel}/\rho^0_{\parallel} \propto \sqrt{B}  \sum_{r=1}^{\infty}  \;\f{R_D^r R_T^r}{r^{1/2}}\cos(r\lambda)\, {\cos\left[ r\left(2\pi \f{F_0}{B}+\pi\right) +\phi_{LK}\right]}.
		\la{sdhformula}
		\end{equation}
  The oscillation amplitude can be smeared due to the finite quasiparticle lifetime $\tau$ and nonzero temperature $T$, which result in the well-known Dingle amplitude factor and thermal damping factor, respectively:
		\begin{equation}
		R_D^r:=\exp\bigg({-}\f{r}{2}\f{h}{\var_c\tau_q}\bigg),\as R_T^r:=\f{a_r}{\text{sinh}\,(a_r)},\as a_r:= \f{2\pi^2 r k_BT}{\var_c}, \la{thermalfactor}
	    \end{equation}
with $\var_c=\hbar |eB|/m_c $ the cyclotron energy,  $m_c=(2\pi)^{-1}\partial S/\partial E$ the cyclotron mass, and $\tau_q$ the quantum lifetime. We first determine $m_c$ via the temperature dependence of oscillation amplitude (Fig. S \ref{mc}b), while the exponential growth of oscillation amplitude with increasing magnetic field stands for the enlarging ratio between the cyclotron energy and the Landau level broadening due to scattering events and therefore can be used to obtain the quantum lifetime. The relation between the field dependence of quantum oscillation amplitude and quantum life time $\tau_q$ can expressed as:
		\begin{equation}
		f(B^{-1})=\ln [Amp \sqrt{B} \sinh (14.69 m_c T / B)] = -\frac{\pi m_c}{e \tau_q} B^{-1} +Const. \la{tauq}
	    \end{equation}
 by extracting the linear slope of the $f(B^{-1})$(Fig. S \ref{Dingle}), $\tau_q$ can be obtained. Its inaccuracy is determined by the uncertainty of the linear fitting (see error bars in Fig. 2). The calculation of the quantum mean free path, $l_q = v_F \tau_q$, further requires the value of Fermi velocity $v_F$, which can be directly calculated with the oscillation frequency and cyclotron mass: 
 \begin{equation}
    v_F = \frac{\hbar k_F}{m_c} = \frac{\sqrt{2\hbar eF}}{m_c}
	    \end{equation}
 This allows us to determine the temperature dependence of $l_q$ for both the $\gamma$ and $\delta$ pockets as shown in Fig.2.

\subsection{Comparison between $h/e$ oscillations and chiral transport}
As illustrated in Fig.2b, the amplitude of the $h/e$ oscillations is strongly suppressed when the magnetic field is rotated away from the in-plane direction. Such a distinct angular dependence directly corresponds to the magneto-chiral transport as previously observed\cite{Guo2022}. The absolute value of the second harmonic voltage $V_{2\omega}$ signal, which stands for the non-reciprocal transport signature in CsV$_3$Sb$_5$, displays an identical angular dependence with the $h/e$ oscillations (Fig. S \ref{chiral}). This striking correspondence is consistently observed in both devices S2 and S4, suggesting the significant modulation of electronic orders in CsV$_3$Sb$_5$ and the possible role of quantum phase coherence in the appearance of the chiral-magneto transport.

\subsection{Simulation of semiclassical Bloch-Lorentz oscillations}
Here, we present the formalism for the semiclassical $B$-periodic oscillation as shown in Fig.2b. We use the Boltzmann transport equation(BTE) to calculate the magneto-conductivity  $\sigma_{zz}$ along $c$-axis of finite-size quasi-two-dimensional materials as shown in (Fig. S \ref{chiral}). 

For quasi-two-dimensional materials, the band dispersion is modeled by 
\begin{equation}
    \varepsilon(\mathbf{k}) = \varepsilon_\parallel(\mathbf{k_\parallel}) - t_z \cos(k_z d),
\end{equation}
where $\mathbf{k}_\parallel = (k_x,k_y)$ is the in-plane momentum and $d$ is the interlayer distance. The out-of-plane coupling $t_z$ is assumed to be weak, i.e., $t_z$ is much smaller than the bandwidth given by $\varepsilon_\parallel$. 

We use linearized ansatz for the distribution function $ f(\mathbf{r},\mathbf{k})$,
\begin{equation}
    f(\mathbf{r},\mathbf{k}) = f^0 - \frac{\partial f^0}{\partial \varepsilon} h(\mathbf{r},\mathbf{k}),
\end{equation} 
where $f^0$ is the Fermi-Dirac distribution function. 

Assuming we impose an electric filed $\mathcal{E}_z$  in the z-direction, the BTE under relaxation time approximation reads 
\begin{equation}
    	\mathbf{v}(\mathbf{k}) \cdot \nabla_\mathbf{r} h - \frac{e}{\hbar}[\mathbf{v}(\mathbf{k}) \times \mathbf{B}]
	\cdot \nabla_\mathbf{k} h - e v_z(k_z) \mathcal{E}_z = -\frac{h(\mathbf{r},\mathbf{k})}{\tau},
\end{equation}
where $\mathbf{v}(k) = \partial \varepsilon(\mathbf{k})/\partial \mathbf{k}$ is Fermi velocity vector and $\mathbf{B} \equiv B (\cos \theta \cos \varphi, \cos \theta \sin \varphi, \sin \theta)$ is external magnetic field. 

For a finite-size sample with length $L$ and width $w$, we impose a completely diffusive boundary condition,
\begin{equation}
    h(\mathbf{r}_B, \mathbf{k}_\parallel, k_z) = 0 \qquad  \mathbf{v}(\mathbf{k}_\parallel) \cdot \hat{\mathbf{n}}_B<0,
\end{equation}
where $\mathbf{r}_B$ is arbitrary position vector at the boundary and $\hat{\mathbf{n}}_B$ is the norm vector at $\mathbf{r}_B$. 

With this boundary condition, the distribution function $h(\mathbf{r},\mathbf{k})$ can be exactly solved by the method of characteristics\cite{vilkelis2023bloch,wang2024transverse}. Therefore, the conductivity can be calculated as 
 \begin{eqnarray}
        \sigma_{z z} &=& \frac{e}{S_{\parallel}\mathcal{E}_z} \int_{S_{\parallel}}
	d^2 \mathbf{r}_\parallel \int_{\text{BZ}} \delta(\varepsilon- \varepsilon_F) 
	h(\mathbf{r},\mathbf{k})v_z(\mathbf{k}) d^3 k
\end{eqnarray}

For an out-of-plane magnetic field with moderate tilting angle $\theta$ up to $30^\circ$, we use an isotropic dispersion $\varepsilon_\parallel(\mathbf{k}_\parallel) = \mathbf{k_\parallel}^2/2m^*$ to model the in-plane dispersion. The solution gives rise to $B$-periodic oscillation of the conductivity as shown in (Fig. S \ref{Theoryinplane}). The physical reason for this is that when the oscillation period is commensurate with the sample width, all the semiclassical trajectories have a zero net $z$-direction displacement over the time of flight. Resulting in a minimal at every commensurate field \cite{vilkelis2023bloch}. Since a moderate tilting angle $\theta$ does not change the in-plane trajectory drastically, the commensurate condition still holds except for a pre-factor $\cos \theta$ due to projection. Thus, the semiclassical B-period oscillation can be observed over a large window of out-of-plane angle $\varphi$ as shown in Fig. 2b. 

We also present the result of varying $\varphi$ at $\theta = 0^\circ$ for the in-plane magnetic field. We try two different models for the in-plane dispersion $\varepsilon_\parallel$: cylindrical and hexagonal models (Fig. S \ref{Theoryop}). Surprisingly, the oscillation vanishes quickly for a cylindrical Fermi surface when rotating away from the principal axis. On the other hand, the oscillation remains visible when rotating to a larger angle in the case of a hexagonal Fermi surface. Yet, multiple oscillation frequencies coexist due to the three dominant directions of Fermi velocities. Therefore the simulation in both limiting cases cannot capture the experimental signatures even qualitatively, in particular regarding the angular dependence.

\subsection{Consistent switch at 45 degrees}
To investigate the universality of the switching behavior illustrated in Fig. 3, we measured the angular dependence of oscillation periodicity in two microstructures, S1 and S4. Devices S1 and S4 have significantly different dimensions at their cross-sections. Device S1 features a narrow channel, with its width being nearly half its depth and, therefore, distinct oscillation periods and amplitudes with field applied either parallel or perpendicular to the crystalline $a$-direction (Fig. S \ref{InplaneS1S4}a-d). Meanwhile, such a difference is much less significant in S4 due to its comparable width and depth (Fig. S \ref{InplaneS1S4}e-g).  Despite their distinction in dimensions, the periodicity switch at 45 degrees modulo 90 degrees was consistently observed in both samples (Fig. S \ref{InplaneEXD}), emphasizing the irrelevance of the device geometry and its close connection to the correlated electronic order.

\subsection{Temperature dependence across various devices}
As microstructures vary in size, the reduction of $h/e$ oscillation amplitude with increasing temperature behaves differently across various devices. Reducing the device's width shifts the onset of the oscillations, defined by the temperature where the oscillation amplitude gets larger than 5\%  of the amplitude at $T$ = 2~K, up to a higher temperature (Fig. S \ref{Tdep}). By linearly extrapolating to zero width, the onset temperature is determined to be about 33 K, which coincides with the $T'$ as explained. It is much higher than the onset of quantum oscillations in CsV$_3$Sb$_5$ (Fig. S \ref{mc}), which displays no variation across all devices of different sizes.

\clearpage

\renewcommand{\figurename}{Fig. S}
\setcounter{figure}{0} 

\begin{figure}
	\centering
\includegraphics[width = 0.98\linewidth]{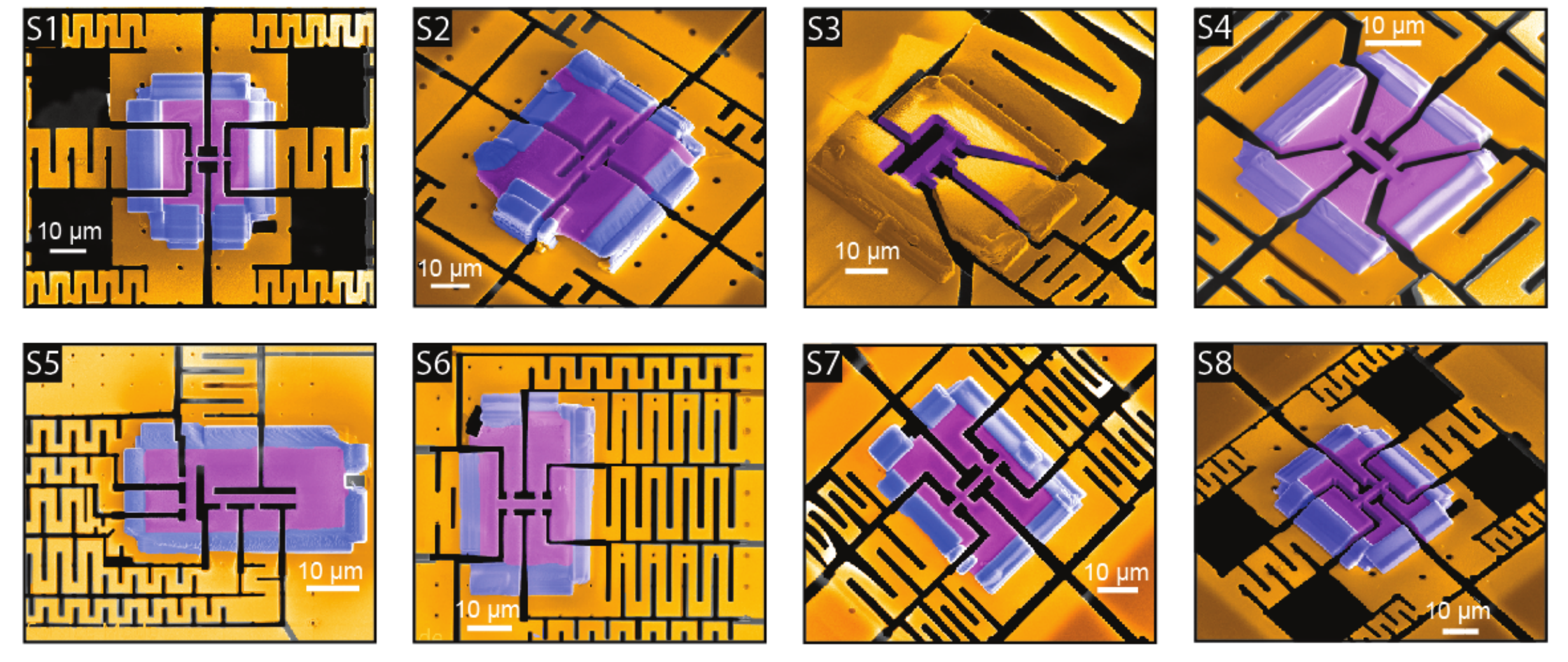}
		\caption{\textbf{Scanning electron microscope (SEM) images of all devices.} All microstructures utilize the soft membrane springs to reduce the thermal differential strain effect at low temperatures.} 
	\label{SEM}
\end{figure}	
\clearpage

\begin{figure}
	\centering
\includegraphics[width = 0.98\linewidth]{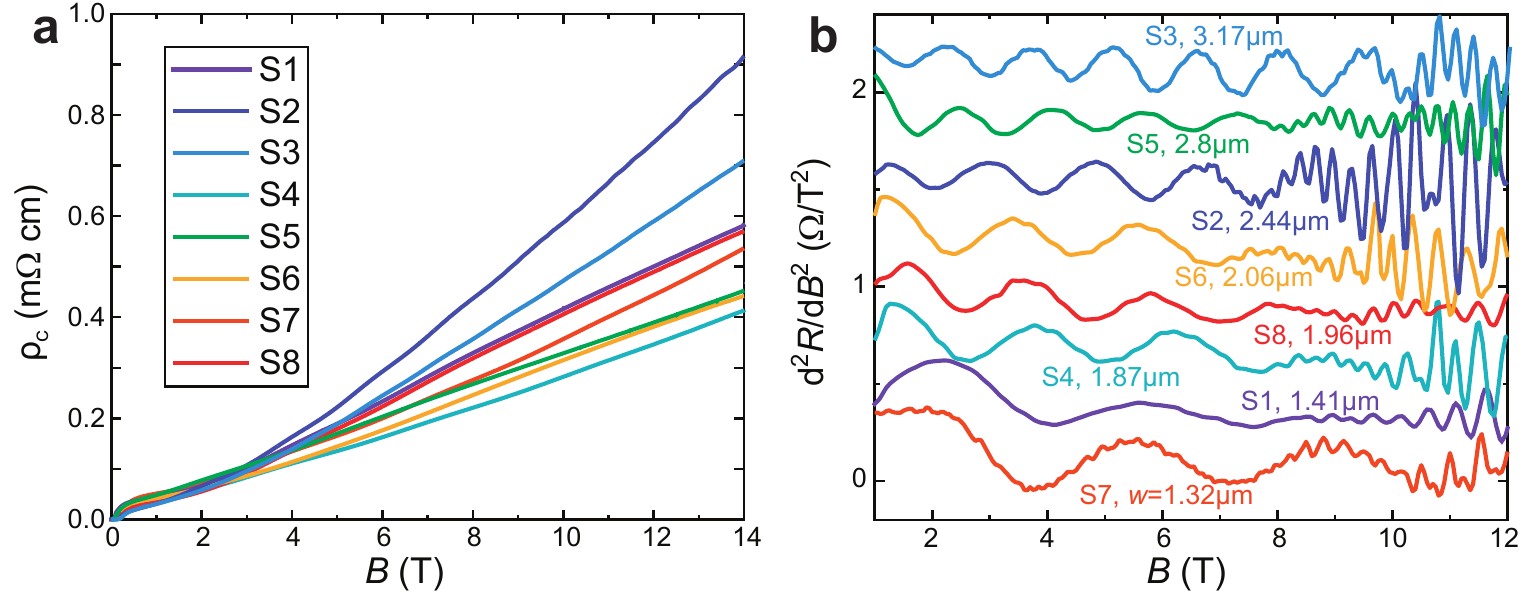}
		\caption{\textbf{Summary of all magnetoresistance and corresponding $h/e$ oscillations.} (a) The magnetoresistance of all devices increases monotonically with increasing field, and the difference between their values is mainly attributed to the inevitable misalignment between the magnetic field and crystalline $a$-direction. (b) The second derivative of the field-dependent magnetoresistance clearly reveals the presence of $h/e$ oscillations in all devices. Note that each curve is shifted by at least 0.3 $\Omega$/T$^2$ for clarity.}
	\label{allSum}
\end{figure}
\clearpage

\begin{figure}
	\centering
\includegraphics[width = 0.98\linewidth]{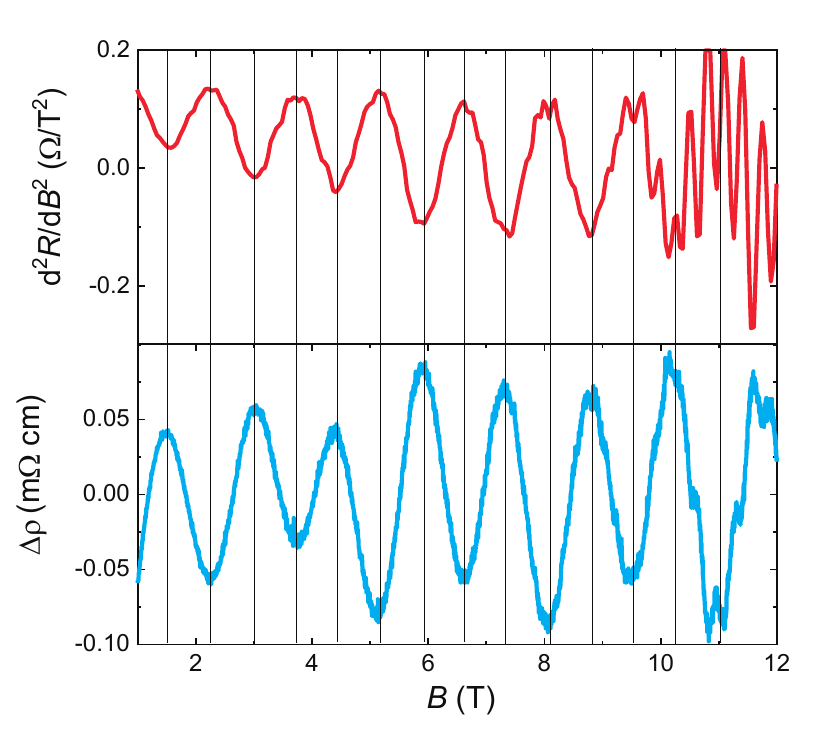}
		\caption{\textbf{Comparison of $h/e$ oscillations subtracted by two different methods.} The $h/e$ oscillations can be clearly resolved by either taking the second field derivative or subtracting the 5th-polynomial fitting as a background. The main difference between these two methods is a $\pi$-phase shift, as demonstrated by the flipped peak-valley correspondence between them.} 
	\label{Compare}
\end{figure}
\clearpage

\begin{figure}
	\centering
\includegraphics[width = 0.98\linewidth]{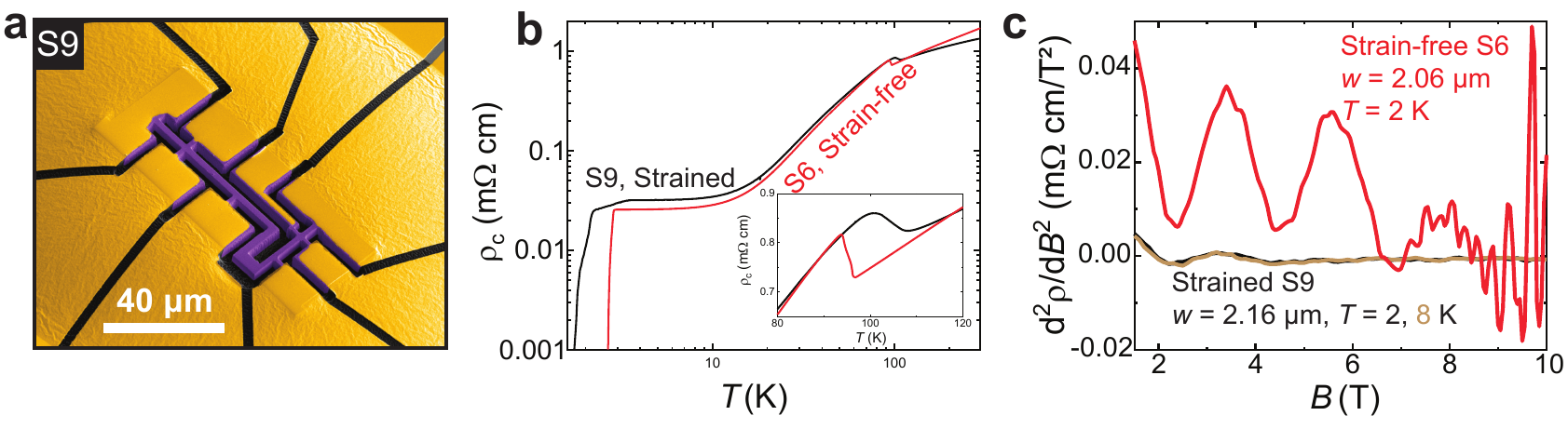}
		\caption{\textbf{Influence of uniaxial strain.} (a) SEM image of device S9. The device is mechanically attached to a sapphire substrate with via a double-component glue droplet. At low temperatures, the thermal expansion coefficient mismatch between the device and the substrate results in a substantial uniaxial strain across the device. (b) $T$-dependent resistivity of strained and nearly strain-free devices. Due to the presence of tensile strain, the charge order temperature is clearly increased, and its broadening is possibly attributed to the strain inhomogeneity. (c) Comparison of $h/e$ oscillations in both devices. The oscillations are strongly suppressed in the strained device and only one oscillation period is visible. This is expected given the extraordinarily strain-sensitivity of CsV$_3$Sb$_5$ as consistently demonstrated in previous works\cite{guo2024correlated,Guo2022}.} 
	\label{Strain}
\end{figure}
\clearpage

\begin{figure}
	\centering
\includegraphics[width = 0.98\linewidth]{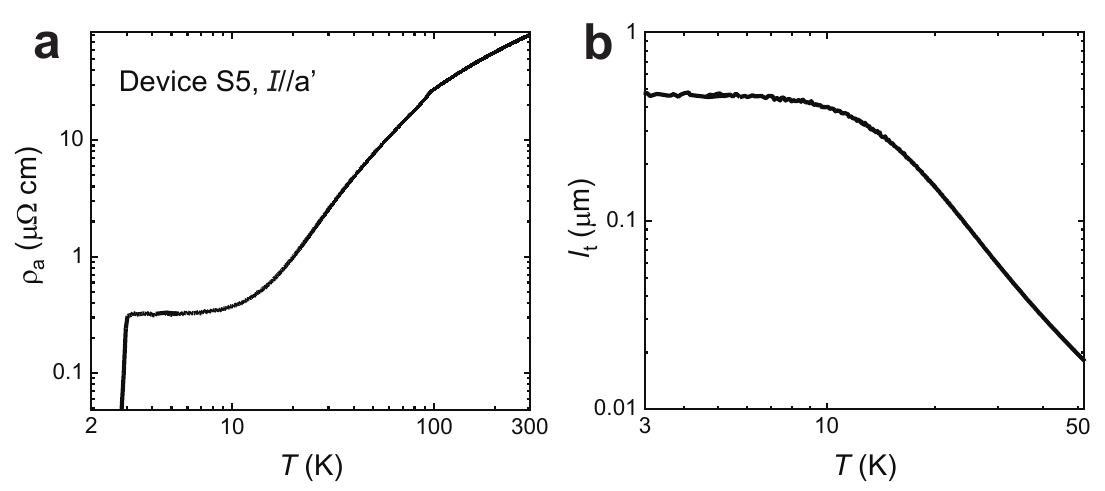}
		\caption{\textbf{Transport mean free path extracted from Drude model.} (a) In-plane resistivity measured with current applied perpendicular to the $a$-direction. The kink at 94 K stands for the charge-density-wave transition, while the superconducting transition is at $T_{\text{c}}$ = 2.8 K. Both values agree well with the previous report on bulk samples\cite{DresdenThermal,ortiz2020cs}, demonstrating the unchanged device properties after FIB-fabrication. (b) Transport mean free path estimated from the in-plane resistivity based on the Drude model.} 
	\label{Drude}
\end{figure}
\clearpage

\begin{figure}
	\centering
\includegraphics[width = 0.98\linewidth]{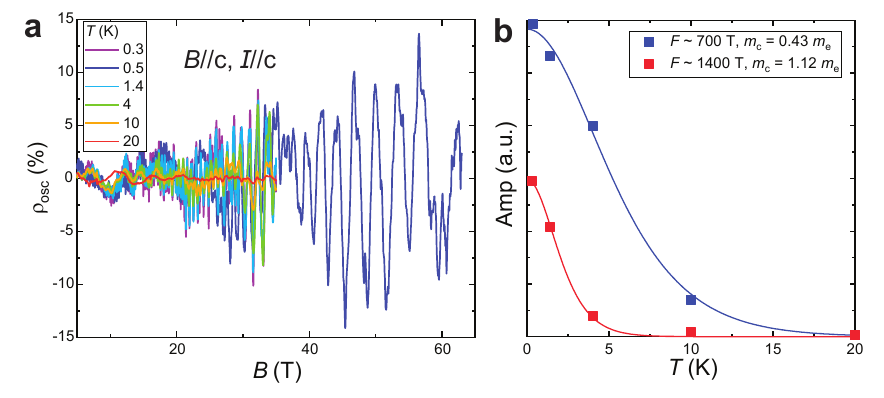}
		\caption{\textbf{Quantum oscillations and cyclotron mass analysis.} (a) Quantum oscillations at various temperatures. The oscillations are obtained from the magnetoresistance by subtracting a 5th-order polynomial fitting as a background. The measurement at $T$ = 0.5 K is performed in a pulsed magnet up to 63 T, while others are measured in a water-cooled static magnet up to 35 T. (b) Lifshitz-Kosevich fitting to the temperature dependence of oscillation amplitudes determines the cyclotron masses of two main Fermi pockets in CsV$_3$Sb$_5$.} 
	\label{mc}
\end{figure}
\clearpage

\begin{figure}
	\centering
\includegraphics[width = 0.98\linewidth]{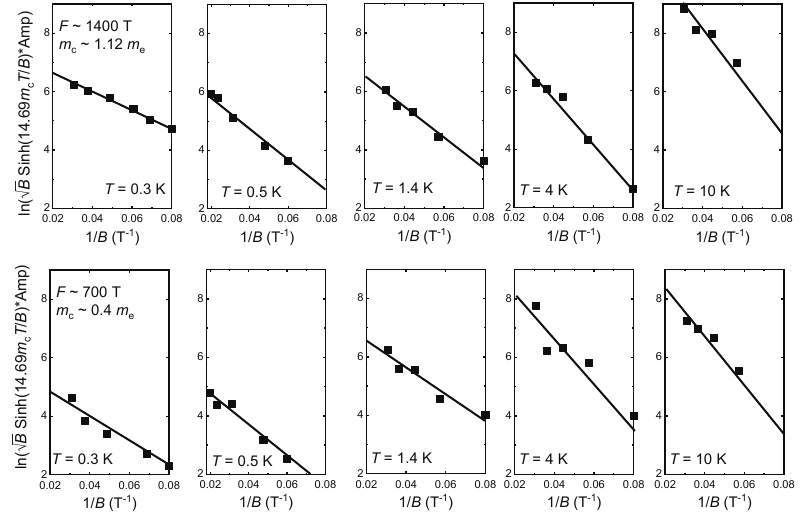}
		\caption{\textbf{Quantum mean free path extracted from Dingle analysis.} Extrapolation of field-dependent oscillation amplitude yields the quantum lifetime for each Fermi surface at various temperatures. The linear fitting coefficient leads to the determination of the quantum mean free path as presented in Fig. 2a.} 
	\label{Dingle}
\end{figure} 
\clearpage

\begin{figure}
	\centering
\includegraphics[width = 0.98\linewidth]{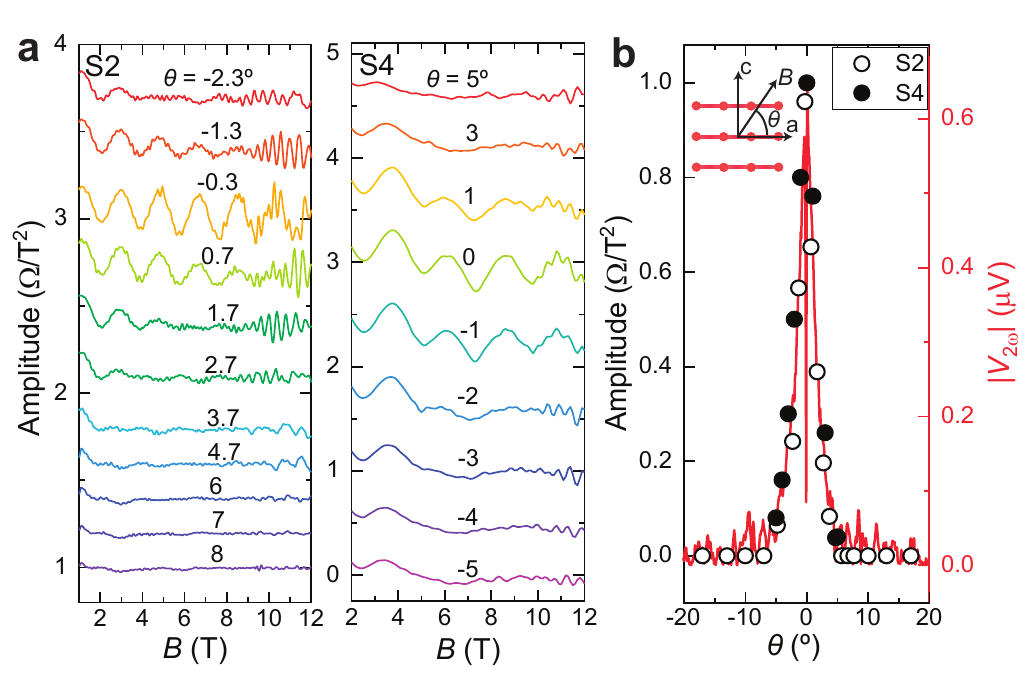}
		\caption{\textbf{Identical angular dependence of chiral magneto transport and $h/e$ oscillations.} (a) $h/e$ oscillations measured at different out-of-plane angle $\theta$ for both device S2 and S4. The oscillation becomes almost invisible at angles larger than 5$^{\circ}$. (b) Comparison between $h/e$ oscillations and the non-reciprocal transport signature $V_{2\omega}$\cite{Guo2022}. Their identical angular dependence suggests the close relation between them.} 
	\label{chiral}
\end{figure}
\clearpage

\begin{figure}
	\centering
\includegraphics[width = 0.98\linewidth]{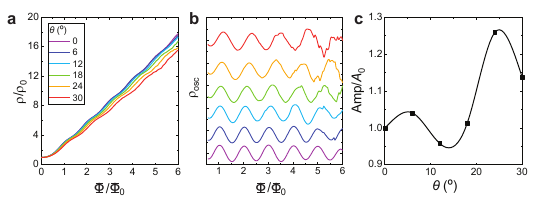}
		\caption{\textbf{Angular dependence of simulated Bloch-Lorentz oscillations based on a cylindrical Fermi surface.} (a) and (b) stands for the simulated magnetoresistance and Bloch-Lorentz oscillations, respectively. With the magnetic field rotated from in-plane to out-of-plane direction, the oscillation amplitude merely changes slightly, as shown in (c), which is distinct from the experimental observations.} 
	\label{Theoryinplane}
\end{figure}
\clearpage

\begin{figure}
	\centering
\includegraphics[width = 0.98\linewidth]{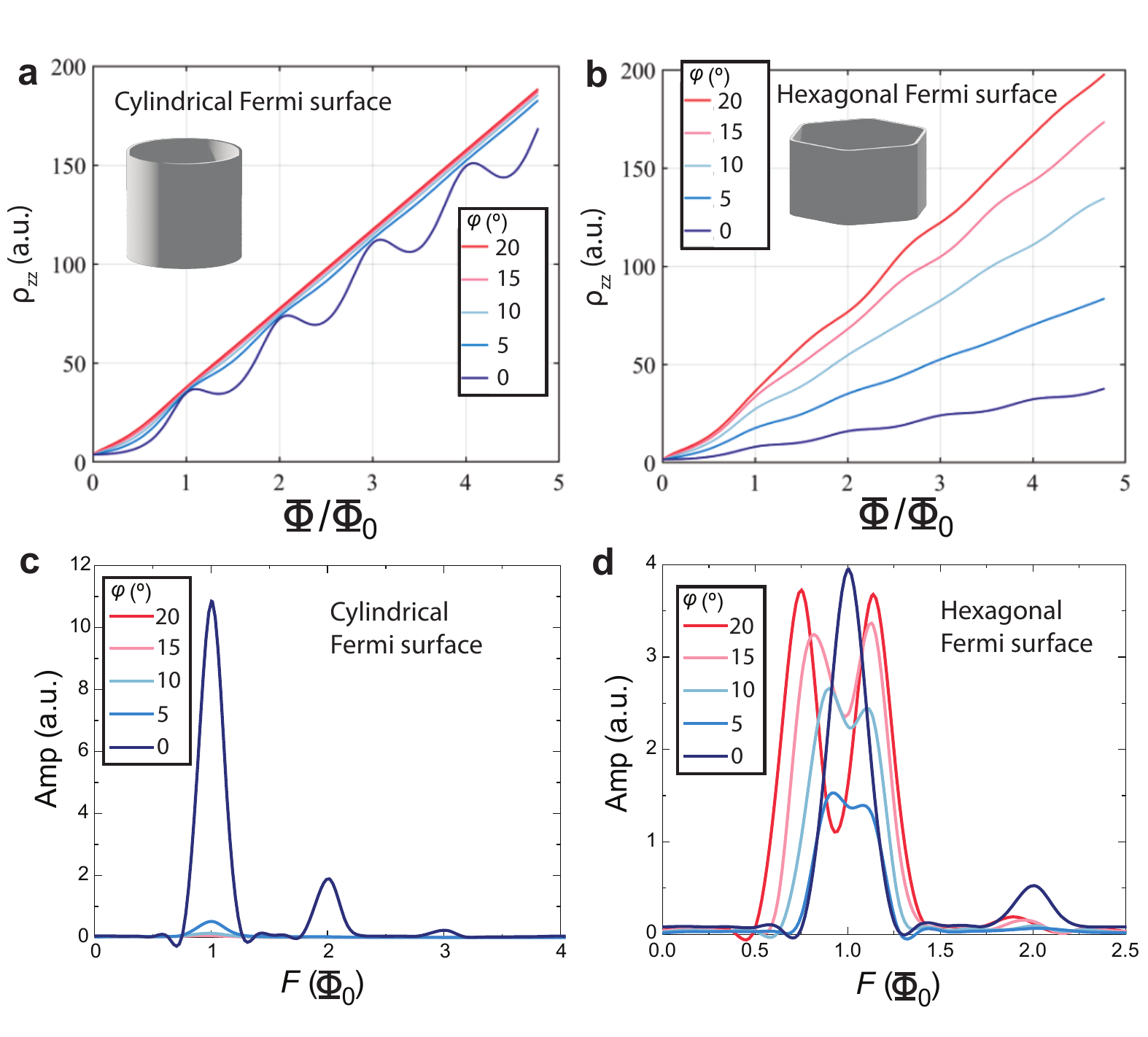}
		\caption{\textbf{Comparison between cylindrical and hexagonal Fermi surface.} (a) and (b) stands for the Bloch-Lorentz oscillation simulation for cylindrical and hexagonal Fermi surfaces respectively, while (c) and (d) present the FFT analysis of the oscillatory parts in both cases. With the magnetic field rotated in-plane, the oscillation amplitude varies quickly in the case of cylindrical Fermi surface, while its change in the hexagonal case is not as significant.} 
	\label{Theoryop}
\end{figure}
\clearpage

\begin{figure}
	\centering
\includegraphics[width = 0.98\linewidth]{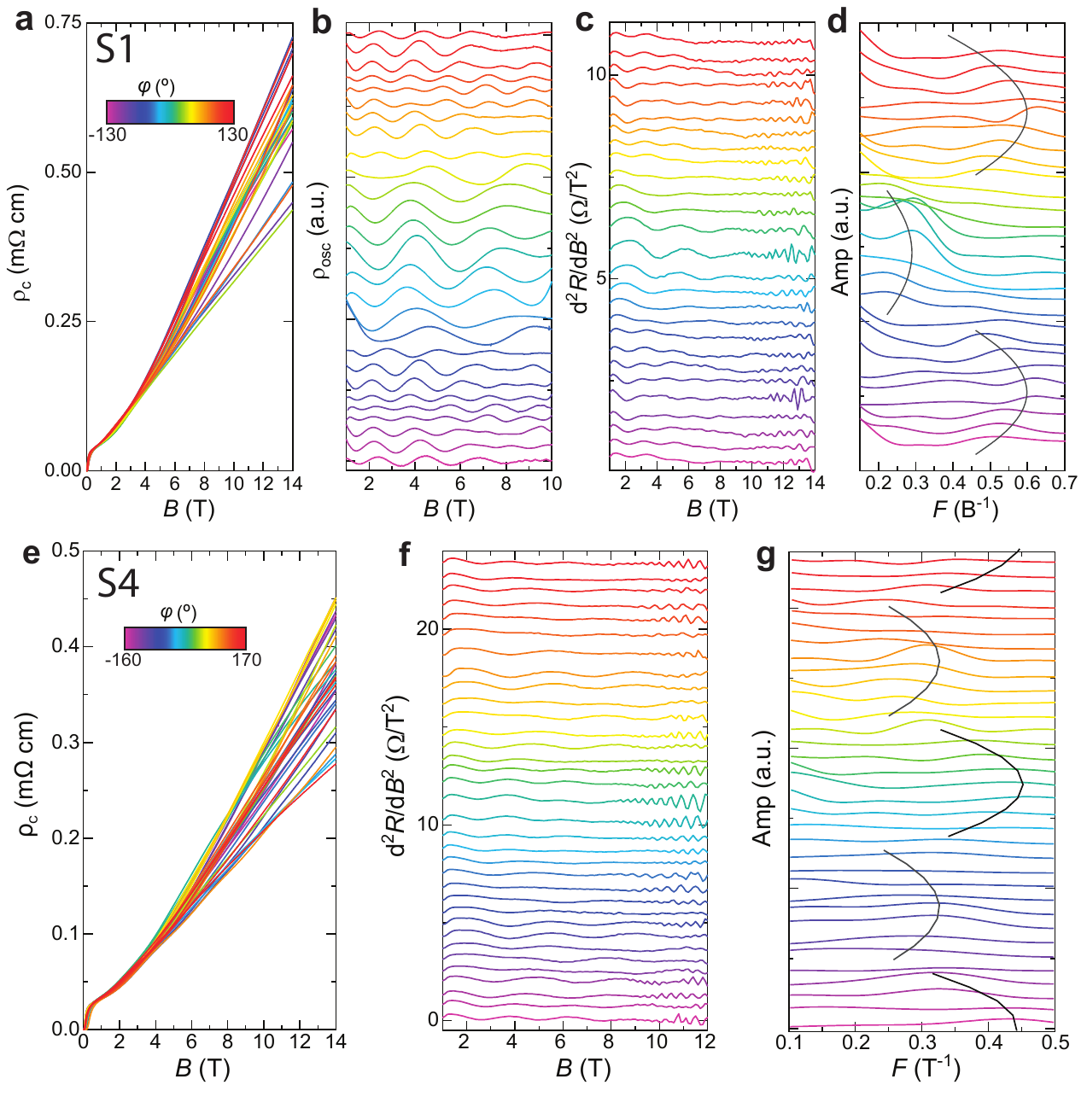}
		\caption{\textbf{In-plane angle-dependent  $h/e$ oscillations of S1 and S4.} (a) Angular dependence of magnetoresistance up to 14T. The measurements are conducted with 10$^{\circ}$ steps from -130$^{\circ}$ to 130$^{\circ}$. (b) and (c) display the $h/e$ oscillations extracted from polynomial background subtraction and second derivative, respectively. (d) FFT spectrum of $h/e$ oscillations at different angles. The black line indicates the guideline for identified oscillation frequency peaks. (e) Angle-dependent magnetoresistance of S4 measured with a rotation from -160$^{\circ}$ to 170$^{\circ}$ in 10$^{\circ}$ steps. (f) second derivative of magnetoresistance at all angles. (g) FFT spectrum of the oscillations at various angles.} 
	\label{InplaneS1S4}
\end{figure}
\clearpage

\begin{figure}
	\centering
\includegraphics[width = 0.98\linewidth]{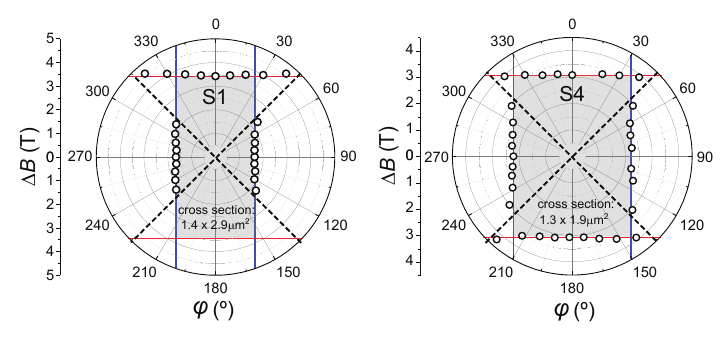}
		\caption{\textbf{Angular dependence of oscillation periods in device S1 and S4.} Polar plot of oscillation period consistently demonstrates the switching behavior at 45$^{\circ}$ modulo 90$^{\circ}$. The consistent observation of such behavior in two devices of distinct geometries further emphasizes that the switching has no correspondence to the sample geometry. }
	\label{InplaneEXD}
\end{figure}
\clearpage

\begin{figure}
	\centering
\includegraphics[width = 0.98\linewidth]{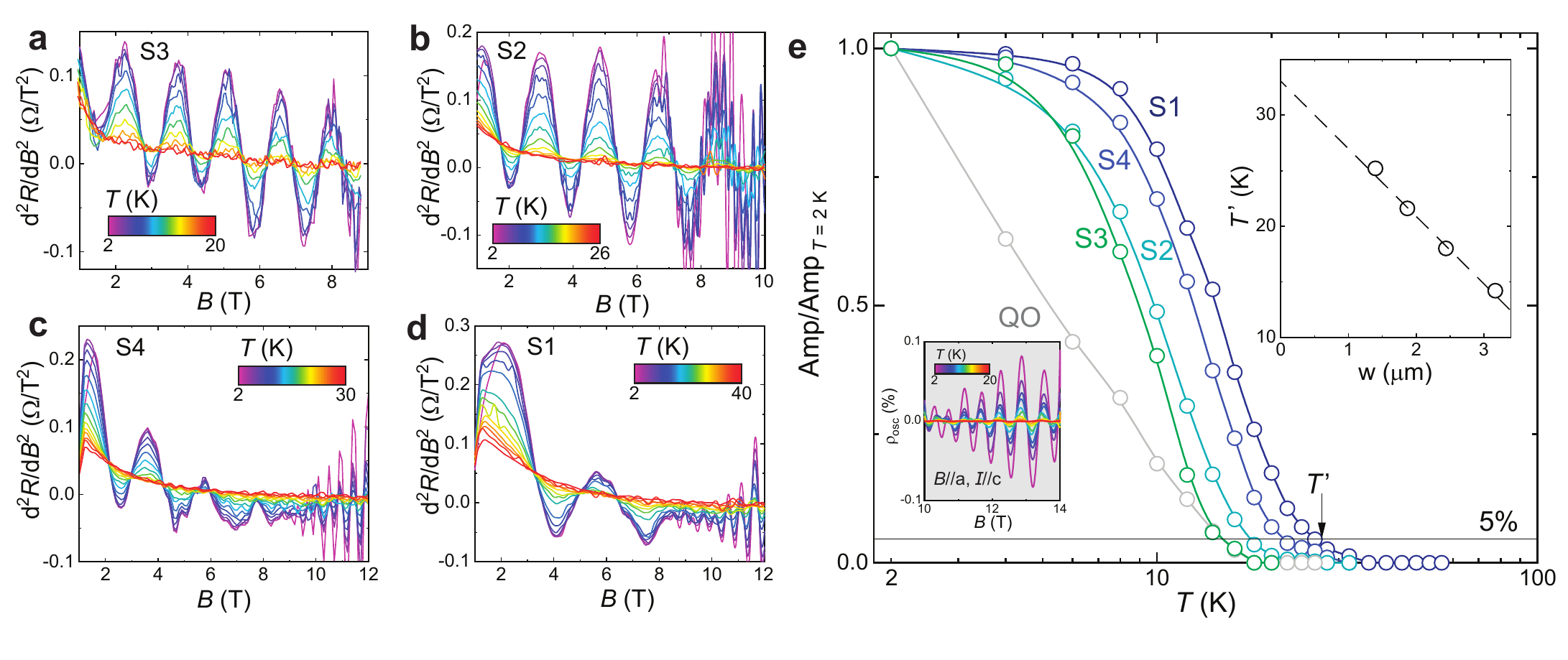}
		\caption{\textbf{Temperature dependence of $h/e$ oscillation amplitude.} (a)-(d) display the temperature dependence of $h/e$ oscillations in S3, S2, S4, S1, respectively. All measurements are conducted from 2~K to elevated temperatures in 2~K steps. (e) Temperature dependencies of oscillation amplitude across various devices and their comparison to quantum oscillations. The left-hand inset presents the temperature-dependent quantum oscillations measured with current and magnetic fields along the $c$- and $a$-direction. The right-hand inset displays the interpolation of $T'$ at zero sample width by linearly extrapolating the temperature where the oscillation amplitude reaches 5\% of the value at 2~K. This yields a $T$' $\approx$ 33 K, which is almost identical to $T$' determined via other experimental results\cite{Guo2022,guo2024correlated}.}
	\label{Tdep}
\end{figure}

\clearpage